\documentclass{article}%
\usepackage{amsmath}
\usepackage{amssymb}
\usepackage{hyperref}
\usepackage{color}
\usepackage{fancybox}
\usepackage{fix2col}
\usepackage{amsfonts}
\usepackage{graphicx}%
\providecommand{\U}[1]{\protect\rule{.1in}{.1in}}
\newtheorem{theorem}{Theorem}

\newtheorem{exercise}[theorem]{Exercise}

\begin{document}

\title{\emph{Quantum Mechanics in P}$\hbar\hspace{-0.02in}$\emph{ase
Space\marginpar{ANL-HEP-PR-11-31 and UMTG-22}}}
\author{Thomas L. Curtright$^{\S }$ and Cosmas K. Zachos$^{\sharp}\bigskip$\\$^{\S }$Department of Physics, University of Miami\\Coral Gables, FL 33124-8046, USA\\$^{\sharp}$High Energy Physics Division, Argonne National Laboratory\\Argonne, IL 60439-4815, USA}
\date{}
\maketitle

\begin{abstract}
Ever since Werner Heisenberg's 1927 paper on uncertainty, there has been
considerable hesitancy in simultaneously considering positions and momenta in
quantum contexts, since these are incompatible observables. \ But this
persistent discomfort with addressing positions and momenta jointly in the
quantum world is not really warranted, as was first fully appreciated by
Hilbrand Groenewold and Jos\'{e} Moyal in the 1940s. \ While the formalism for
quantum mechanics in phase space was wholly cast at that time, it was not
completely understood nor widely known --- much less generally accepted ---
until the late 20th century.$\bigskip$

\end{abstract}

\vfill

When Feynman first unlocked the secrets of the path integral formalism and
presented them to the world, he was publicly rebuked \cite{Gleick}:
\textquotedblleft It was obvious, Bohr said, that such trajectories violated
the uncertainty principle.\textquotedblright\ \ 

However, in this case,\footnote{Unlike
\href{http://en.wikipedia.org/wiki/Bohr-Einstein_debates}{the more famous
cases} where Bohr criticised thought experiments proposed by Einstein, at the
1927 and 1930 Solvay Conferences.} Bohr was wrong. \ Today path integrals are
universally recognized and widely used as an alternative framework to describe
quantum behavior, equivalent to although conceptually distinct from the usual
Hilbert space framework, and therefore completely in accord with Heisenberg's
uncertainty principle. \ The different points of view offered by the Hilbert
space and path integral frameworks combine to provide greater insight and
depth of understanding.

Similarly, many physicists\marginpar{%
%TCIMACRO{\FRAME{itbpFU}{1.1831in}{1.4944in}{0in}{\Qcb{R Feynman}}%
%{}{feynman.eps}{\special{ language "Scientific Word";  type "GRAPHIC";
%width 1.1831in;  height 1.4944in;  depth 0in;  original-width 3.0675in;
%original-height 3.8943in;  cropleft "0";  croptop "1";  cropright "1";
%cropbottom "0";  filename 'Feynman.eps';file-properties "XNPEU";}} }%
%BeginExpansion
{\parbox[b]{1.1831in}{\begin{center}
\includegraphics[
height=1.4944in,
width=1.1831in
]%
{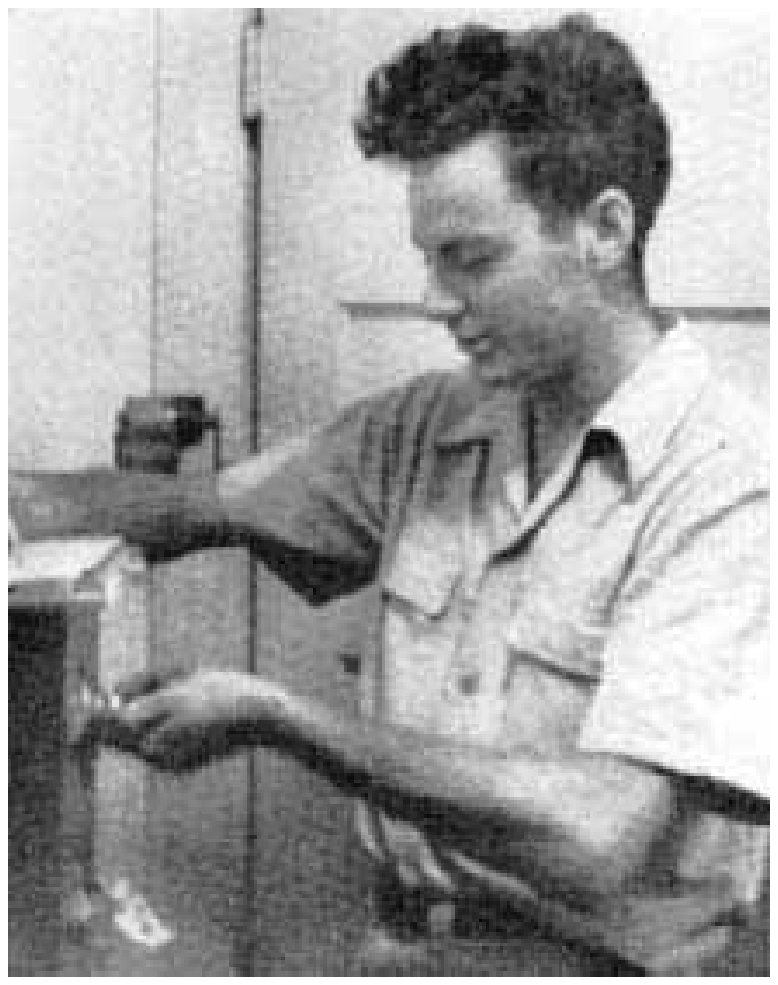}%
\\
R Feynman
\end{center}}}
%EndExpansion
\par%
%TCIMACRO{\FRAME{itbpFU}{1.0577in}{1.4935in}{0in}{\Qcb{N Bohr}}{}%
%{bohr.eps}{\special{ language "Scientific Word";  type "GRAPHIC";
%width 1.0577in;  height 1.4935in;  depth 0in;  original-width 3.1938in;
%original-height 4.5506in;  cropleft "0";  croptop "1";  cropright "1";
%cropbottom "0";  filename 'Bohr.eps';file-properties "XNPEU";}} }%
%BeginExpansion
{\parbox[b]{1.0577in}{\begin{center}
\includegraphics[
height=1.4935in,
width=1.0577in
]%
{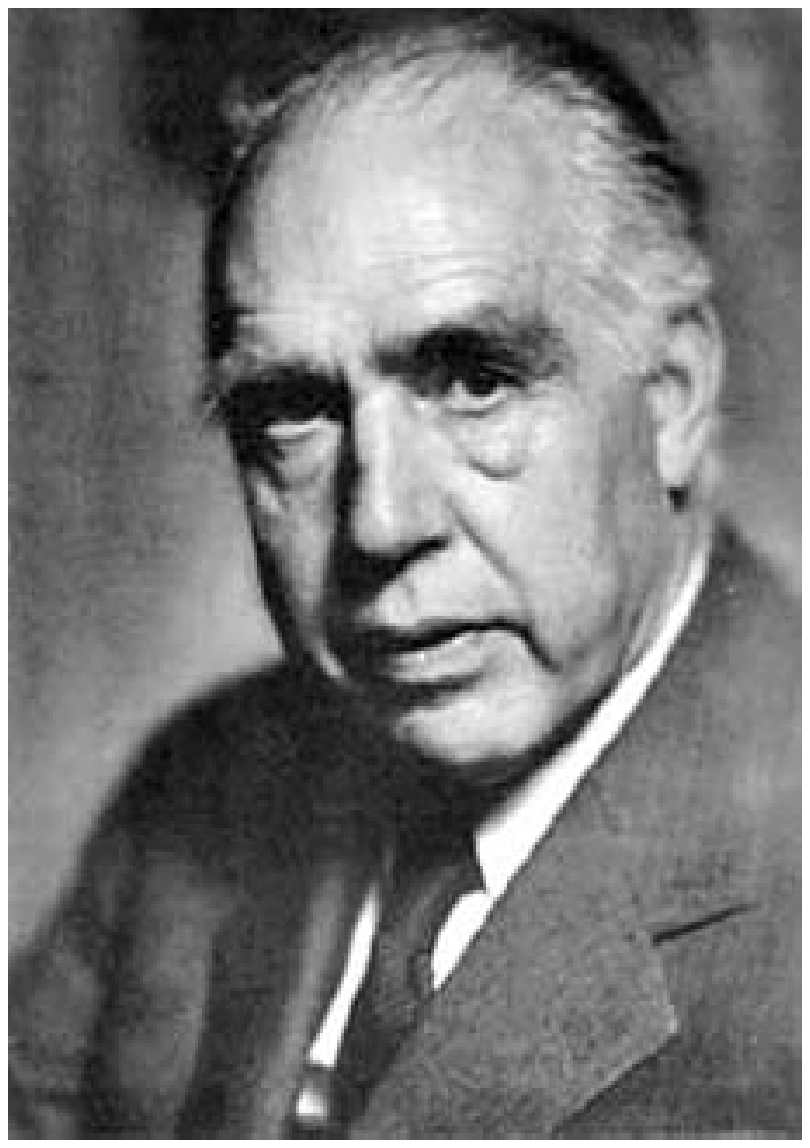}%
\\
N Bohr
\end{center}}}
%EndExpansion
} hold the conviction that classical-valued position and momentum variables
should not be simultaneously employed in any meaningful formula expressing
quantum behavior, simply because this would also seem to violate the
uncertainty principle (see \href{#DiracBox}{Dirac}). \ 

However, they too are wrong. \ Quantum mechanics (QM) \emph{can} be
consistently and autonomously formulated in phase space, with c-number
position and momentum variables simultaneously placed on an equal footing, in
a way that fully respects Heisenberg's principle. \ This other quantum
framework is equivalent to both the Hilbert space approach and the path
integral formulation. \ Quantum mechanics in phase space (QMPS) thereby gives
a third point of view which provides still more insight and understanding.

What follows is the somewhat erratic story of this third formulation.

\vfill\newpage

The foundations of this remarkable picture of quantum mechanics were laid out
by H Weyl and E Wigner around 1930.\marginpar{%
%TCIMACRO{\FRAME{itbpFU}{1.2324in}{1.4935in}{0in}{\Qcb{H Weyl}}{}%
%{weyl.eps}{\special{ language "Scientific Word";  type "GRAPHIC";
%width 1.2324in;  height 1.4935in;  depth 0in;  original-width 3.7291in;
%original-height 4.5394in;  cropleft "0";  croptop "1";  cropright "1";
%cropbottom "0";  filename 'Weyl.eps';file-properties "XNPEU";}} }%
%BeginExpansion
{\parbox[b]{1.2324in}{\begin{center}
\includegraphics[
height=1.4935in,
width=1.2324in
]%
{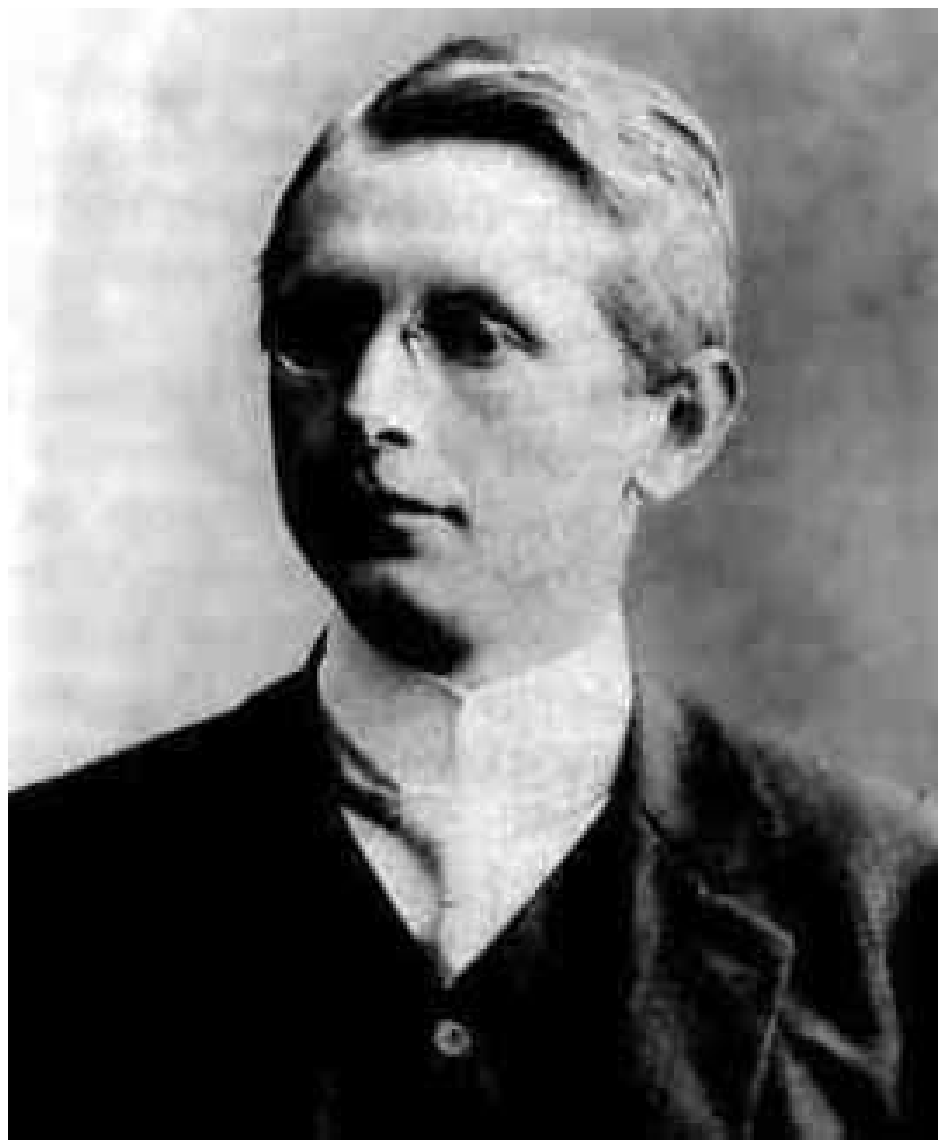}%
\\
H Weyl
\end{center}}}
%EndExpansion
\par%
%TCIMACRO{\FRAME{itbpFU}{1.9856in}{1.4944in}{0in}{\Qcb{W Heisenberg and E
%Wigner}}{}{heisenbergwigner.eps}{\special{ language "Scientific Word";
%type "GRAPHIC";  maintain-aspect-ratio TRUE;  display "USEDEF";
%valid_file "F";  width 1.9856in;  height 1.4944in;  depth 0in;
%original-width 8.3506in;  original-height 6.2578in;  cropleft "0";
%filename 'HeisenbergWigner.eps';file-properties "XNPEU";}} }%
%BeginExpansion
{\parbox[b]{1.9856in}{\begin{center}
\includegraphics[
height=1.4944in,
width=1.9856in
]%
{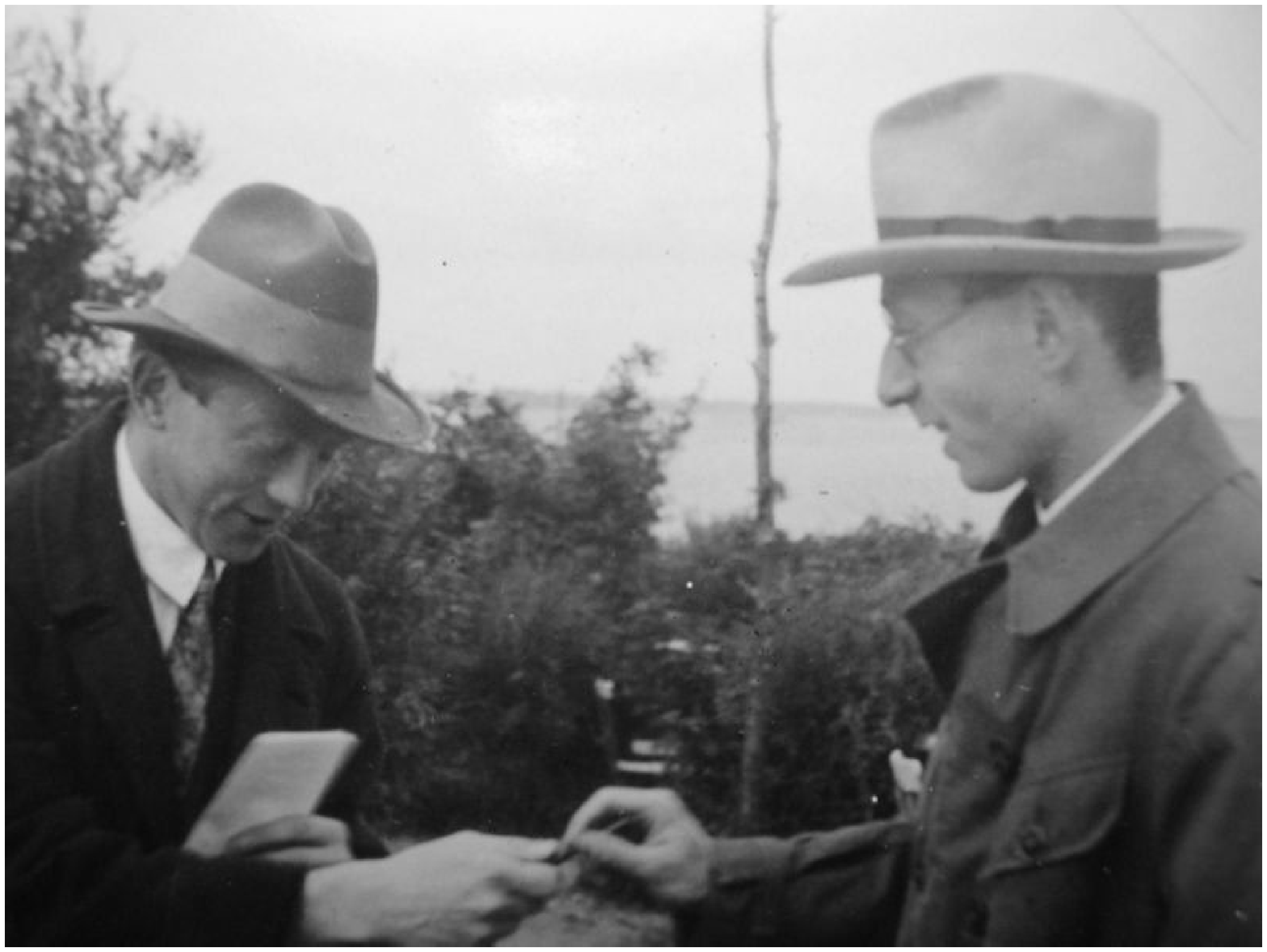}%
\\
W Heisenberg and E Wigner
\end{center}}}
%EndExpansion
} \ But the full, self-standing theory was put together in a crowning
achievement by two unknowns, at the very beginning of their physics careers,
independently of each other, during World War II: \ H Groenewold in Holland
and J Moyal in England (see \href{#GroenewoldBox}{Groenewold} and
\href{#MoyalBox}{Moyal}). \ It was only published after the end of the war,
under not inconsiderable adversity, in the face of opposition by established
physicists; and it took quite some time for this uncommon achievement to be
appreciated and utilized by the community.\footnote{Perhaps this is because it
emerged nearly simultaneously with the path integral and associated
diagrammatic methods of Feynman, whose flamboyant application of those methods
to the field theory problems of the day captured the attention of physicists
worldwide, and thus overshadowed other theoretical developments.}

The net result is that quantum mechanics works smoothly and consistently in
phase space, where position coordinates and momenta blend together closely and
symmetrically. \ Thus, sharing a common arena and language with classical
mechanics \cite{Nolte}, QMPS connects to its classical limit more naturally
and intuitively than in the other two familiar alternate pictures, namely, the
standard formulation through operators in Hilbert space, or the path integral formulation.

Still, as every physics undergraduate learns early on, classical phase space
is built out of \textquotedblleft c-number\textquotedblright\ position
coordinates and momenta, $x$ and $p$, ordinary commuting variables
characterizing physical particles; whereas such observables are usually
represented in quantum theory by operators that do not commute. \ How then can
the two be reconciled? \ The ingenious technical solution to this problem was
provided by Groenewold in 1946, and consists of a special binary operation,
the $\star$-product (see \href{#StarProductsBox}{Star Product}), which enables
$x$ and $p$ to maintain their conventional classical interpretation, but which
also permits $x$ and $p$ to combine more subtly than conventional classical
variables; in fact to combine in a way that is \emph{equivalent} to the
familiar operator algebra of Hilbert space quantum theory.

Nonetheless, expectation values of quantities measured in the lab
(observables) are computed in this picture of quantum mechanics by simply
taking integrals of conventional functions of $x$ and $p$ with a
quasi-probability density in phase space, the Wigner function --- essentially
the density matrix in this picture. \ But, unlike a Liouville probability
density of classical statistical mechanics, this density can take provocative
negative values and, indeed, these can be reconstructed from lab measurements
\cite{Leibfried}.

How does one interpret these \textquotedblleft negative
probabilities\textquotedblright\ in phase space? \ The answer is that, like a
magical invisible mantle, the uncertainty principle manifests itself in this
picture in unexpected but quite powerful ways, and prevents the formulation of
unphysical questions, let alone paradoxical answers (see
\href{#UncertaintyPrincipleBox}{Uncertainty Principle}).

Remarkably, the phase-space formulation was reached from rather different,
indeed, apparently unrelated, directions. \ To the extent this story has a
beginning, this may well have been H Weyl's remarkably rich 1927 paper
\cite{Weyl} (reprinted in \cite{QMPS}) shortly after the triumphant
formulation of conventional QM. \ This paper introduced the correspondence of
phase-space functions to \textquotedblleft Weyl-ordered" operators in Hilbert
space. \ It relied on a systematic, completely symmetrized ordering scheme of
noncommuting operators $\boldsymbol{X}$ and $\boldsymbol{P}$.

Eventually it would become apparent that this was a mere change of
representation. \ But as expressed in his paper at the time \cite{Weyl}, Weyl
believed that this map, which now bears his name, is \textquotedblleft
the\textquotedblright\ quantization prescription ---\ superior to other
prescriptions --- the elusive bridge extending classical mechanics to the
operators of the broader quantum theory containing it; effectively, then, some
extraordinary \textquotedblleft right way\textquotedblright\ to a
\textquotedblleft correct\textquotedblright\ quantum theory.\newpage

However, Weyl's correspondence \emph{fails} to transform the square of the
classical angular momentum to its accepted quantum analog; and therefore it
was soon recognized to be an elegant, but not intrinsically special
quantization prescription. \ As physicists slowly became familiar with the
existence of different quantum systems sharing a common classical limit, the
quest for the right way\ to quantization was partially mooted.

In 1931, in establishing the essential uniqueness of Schr\"{o}dinger's
representation in Hilbert space, von Neumann utilized the Weyl correspondence
as an equivalent abstract representation of the Heisenberg group in the
Hilbert space operator formulation. \ For completeness' sake, ever the curious
mathematician's foible, he worked out the analog (isomorph) of operator
multiplication in phase space. \ He thus effectively discovered the
convolution rule governing the noncommutative composition of the corresponding
phase-space functions --- an early version of the $\star$-product.

Nevertheless, possibly because he did not use it for anything at the time, von
Neumann oddly ignored his own early result on the $\star$-product and just
proceeded to postulate correspondence rules between classical and quantum
mechanics in his very influential 1932 book on the foundations of QM
\cite{Neumann}. \ In fact, his ardent follower, Groenewold, would use the
$\star$-product to show some of the expectations formed by these rules to be
untenable, 15 years later. \ But we are getting ahead of the story.\marginpar{%
%TCIMACRO{\FRAME{itbpFU}{1.1113in}{1.4935in}{0in}{\Qcb{J von Neumann}}%
%{}{vonneumann.eps}{\special{ language "Scientific Word";  type "GRAPHIC";
%width 1.1113in;  height 1.4935in;  depth 0in;  original-width 3.3624in;
%original-height 4.5506in;  cropleft "0";  croptop "1";  cropright "1";
%cropbottom "0";  filename 'VonNeumann.eps';file-properties "XNPEU";}} }%
%BeginExpansion
{\parbox[b]{1.1113in}{\begin{center}
\includegraphics[
height=1.4935in,
width=1.1113in
]%
{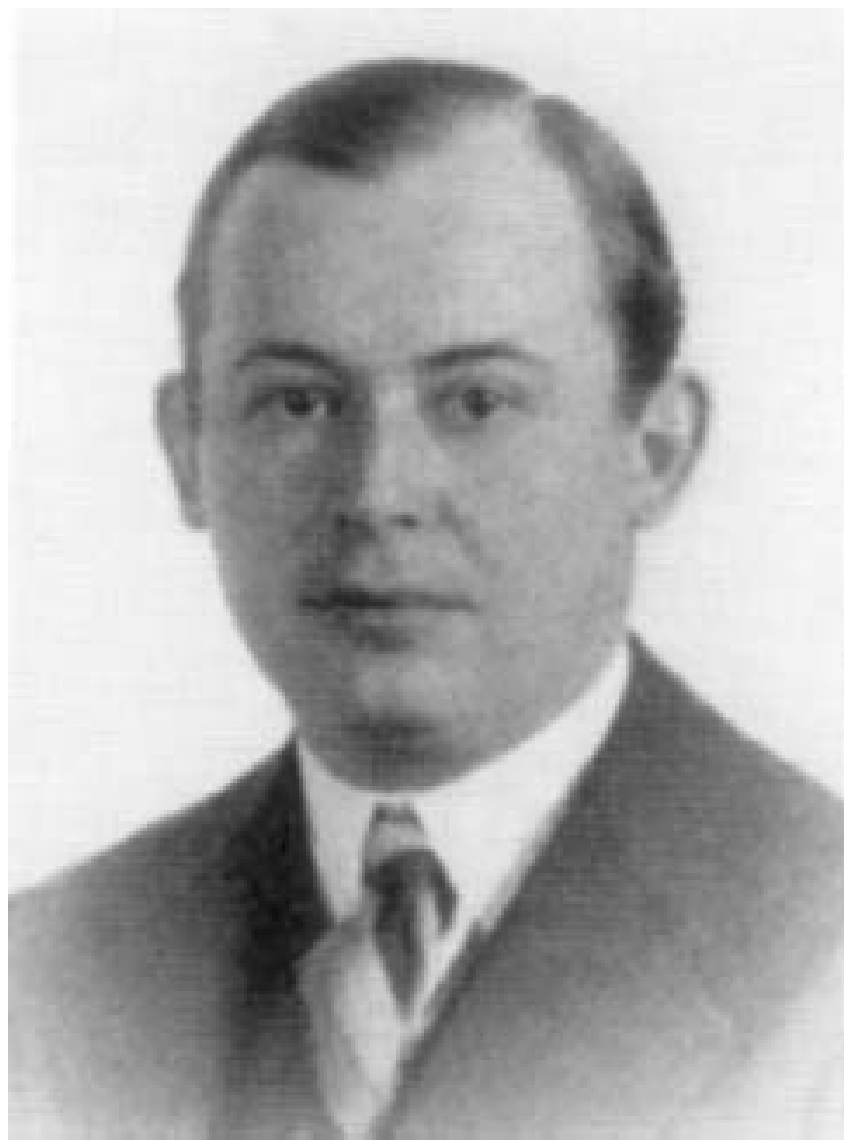}%
\\
J von Neumann
\end{center}}}
%EndExpansion
}

Very soon after von Neumann's paper appeared, in 1932, Eugene Wigner
approached the problem from a completely different point of view, in an effort
to calculate quantum corrections to classical thermodynamic (Boltzmann)
averages. \ Without connecting it to the Weyl correspondence, Wigner
introduced his eponymous function (see \href{#WFBox}{Wigner Functions}), a
distribution which controls quantum-mechanical diffusive flow in phase space,
and thus specifies quantum corrections to the Liouville density of classical
statistical mechanics.

As Groenewold and Moyal would find out much later, it turns out that this WF
maps to the density matrix (up to multiplicative factors of $\hbar$) under the
Weyl map. \ Thus, without expressing awareness of it, Wigner had introduced an
explicit illustration of the inverse map to the Weyl map, now known as the
Wigner map.

Wigner also noticed the WF would assume negative values, which complicated its
conventional interpretation as a probability density function. \ However ---
perhaps unlike his sister's husband --- in time Wigner grew to appreciate that
the negative values of his function were an asset, and not a liability, in
ensuring the orthogonality properties of the formulation's building blocks,
the \textquotedblleft stargenfunctions\textquotedblright\ (see
\href{#SHOBox}{Simple Harmonic Oscillator}).

Wigner further worked out the dynamical evolution law of the WF, which
exhibited the nonlocal convolution features of $\star$-product operations, and
violations of Liouville's theorem. \ But, perhaps motivated by practical
considerations, he did not pursue the formal and physical implications of such
operations, at least not at the time. \ Those and other decisive steps in the
formulation were taken by two young novices, independently, during World War II.\ \ 

In 1946, based on his wartime PhD thesis work, much of it carried out in
hiding, Hip Groenewold published a decisive paper, in which he explored the
consistency of the classical-quantum correspondences envisioned by von
Neumann. \ His tool was a fully mastered formulation of the Weyl
correspondence as an invertible transform, rather than as a consistent
quantization rule. \ The crux of this isomorphism is the celebrated $\star
$-product in its modern form.

Use of this product helped Groenewold demonstrate how Poisson brackets
contrast crucially to quantum commutators (\textquotedblleft Groenewold's
Theorem\textquotedblright). \ In effect, the Wigner map of quantum commutators
is a generalization of Poisson brackets, today called Moyal brackets (perhaps
unfairly, given that Groenewold's work appeared first), which contains Poisson
brackets as their classical limit (technically, a Wigner-Inon\"{u} Lie-algebra
contraction). \ By way of illustration, Groenewold further worked out the
harmonic oscillator WFs. \ Remarkably, the basic polynomials involved turned
out to be those of Laguerre, and not the Hermite polynomials utilized in the
standard Schr\"{o}dinger formulation! \ Groenewold had crossed over to a
different continent.

At the very same time, in England, Joe Moyal was developing effectively the
same theory from a yet different point of view, landing at virtually the
opposite coast of the same continent. \ He argued with Dirac on its validity
(see \href{#DiracBox}{Dirac}) and only succeeded in publishing it, much
delayed, in 1949. \ With his strong statistics background, Moyal focussed on
all expectation values of quantum operator monomials, $\boldsymbol{X}%
^{n}\boldsymbol{P}^{m}$, symmetrized by Weyl ordering, expectations which are
themselves the numerically valued (c-number) building blocks of every quantum
observable measurement.

Moyal saw that these expectation values could be generated out of a
\emph{classical-valued characteristic function in phase space}, which he only
much later identified with the Fourier transform used previously by Wigner.
\ He then appreciated that many familiar operations of standard quantum
mechanics could be apparently bypassed. \ He reassured himself the uncertainty
principle was incorporated in the structure of this characteristic function,
and that it indeed constrained expectation values of \textquotedblleft
incompatible observables.\textquotedblright\ \ He interpreted subtleties in
the diffusion of the probability fluid and the \textquotedblleft negative
probability\textquotedblright\ aspects of it, appreciating that negative
probability is a microscopic phenomenon. \ 

Today, students of QMPS routinely demonstrate as an exercise that, in
$2n$-dimensional phase space, domains where the WF is solidly negative cannot
be significantly larger than the minimum uncertainty volume, $\left(
\hbar/2\right)  ^{n}$, and are thus not amenable to direct observation ---
only indirect inference \cite{Leibfried}.

Less systematically than Groenewold, Moyal also recast the quantum time
evolution of the WF through a deformation of the Poisson bracket into the
Moyal bracket, and thus opened up the way for a direct study of the
semiclassical limit $\hbar\rightarrow0$ as an asymptotic expansion in powers
of $\hbar$ --- \textquotedblleft direct\textquotedblright\ in contrast to the
methods of taking the limit of large occupation numbers, or of computing
expectations of coherent states (see \href{#ClassicalLimitBox}{Classical
Limit}). \ The subsequent applications paper of Moyal with the eminent
statistician Maurice Bartlett also appeared in 1949, almost simulaneously with
Moyal's fundamental general paper. \ There, Moyal and Bartlett calculate
propagators and transition probabilities for oscillators perturbed by
time-dependent potentials, to demonstrate the power of the phase-space
picture.\marginpar{%
%TCIMACRO{\FRAME{itbpFU}{1.1087in}{1.4131in}{0in}{\Qcb{M Bartlett}}%
%{}{bartlett.eps}{\special{ language "Scientific Word";  type "GRAPHIC";
%width 1.1087in;  height 1.4131in;  depth 0in;  original-width 6.2578in;
%original-height 7.9883in;  cropleft "0";  croptop "1";  cropright "1";
%cropbottom "0";  filename 'Bartlett.eps';file-properties "XNPEU";}} }%
%BeginExpansion
{\parbox[b]{1.1087in}{\begin{center}
\includegraphics[
height=1.4131in,
width=1.1087in
]%
{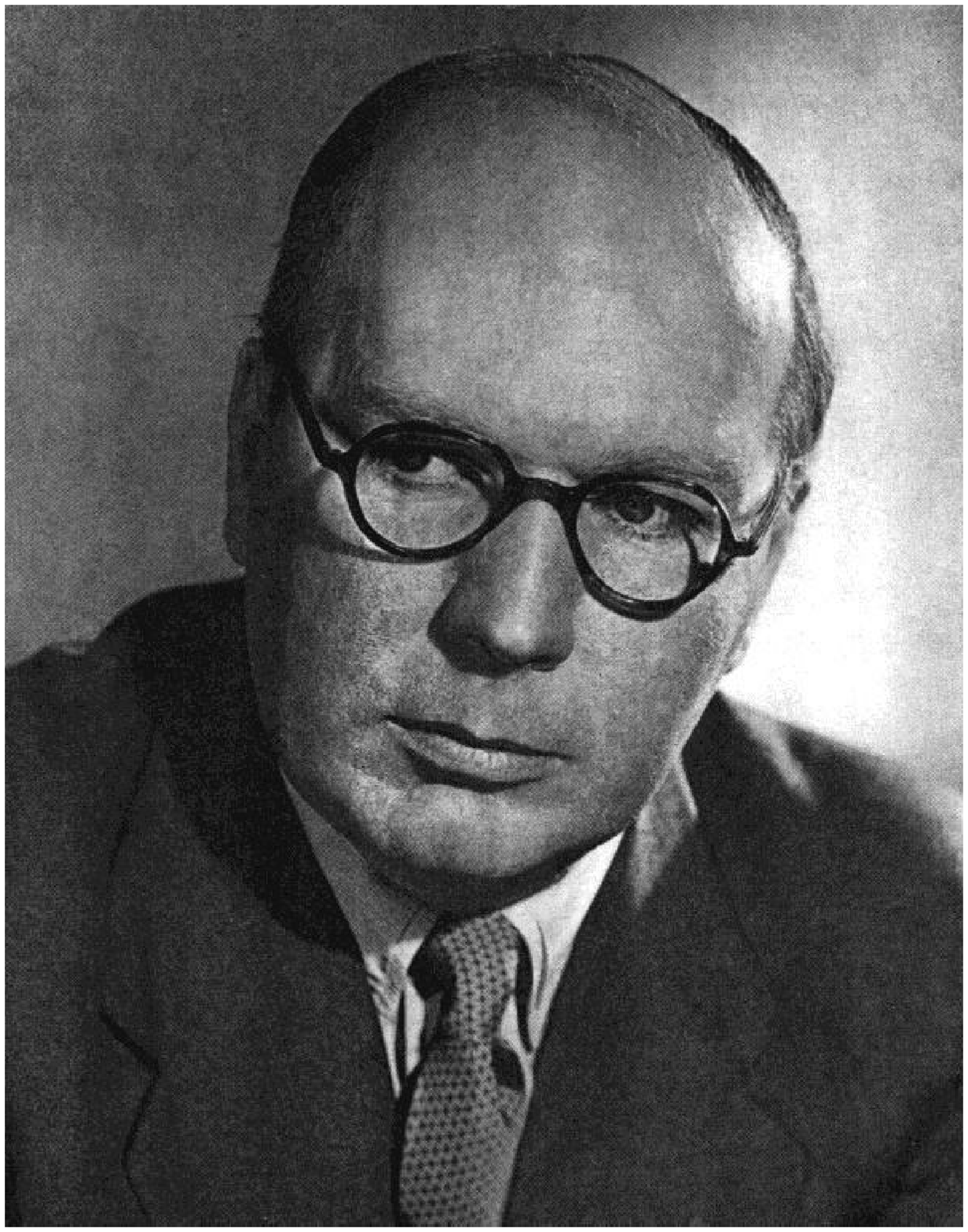}%
\\
M Bartlett
\end{center}}}
%EndExpansion
\par%
%TCIMACRO{\FRAME{itbpFU}{1.0006in}{1.3171in}{0in}{\Qcb{D Fairlie}}%
%{}{fairlie.eps}{\special{ language "Scientific Word";  type "GRAPHIC";
%width 1.0006in;  height 1.3171in;  depth 0in;  original-width 1.4269in;
%original-height 1.8896in;  cropleft "0";  croptop "1";  cropright "1";
%cropbottom "0";  filename 'Fairlie.eps';file-properties "XNPEU";}} }%
%BeginExpansion
{\parbox[b]{1.0006in}{\begin{center}
\includegraphics[
height=1.3171in,
width=1.0006in
]%
{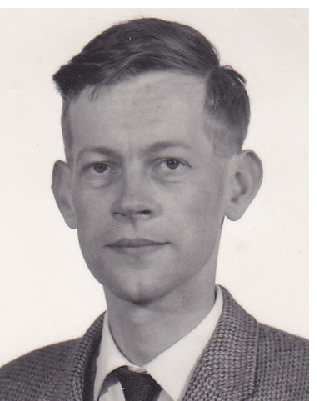}%
\\
D Fairlie
\end{center}}}
%EndExpansion
}

By 1949 the formulation was complete, although few took note of Moyal's and
especially Groenewold's work. \ And in fact, at the end of the war in 1945, a
number of researchers in Paris, such as J Yvon and J Bass, were also
rediscovering the Weyl correspondence and converging towards the same picture,
albeit in smaller, hesitant, discursive, and considerably less explicit steps
(see \cite{QMPS}\ for a guide to this and related literature).

Important additional steps were subsequently carried out by T Takabayasi
(1954), G Baker (1958, his thesis), D Fairlie (1964), and R Kubo (1964)
(again, see \cite{QMPS}).\ These researchers provided imaginative applications
and filled-in the logical autonomy of the picture --- the option, in
principle, to derive the Hilbert-space picture from it, and not vice versa.
\ The completeness and orthogonality structure of the eigenfunctions in
standard QM is paralleled in a delightful shadow-dance, by QMPS $\star
$-operations (see \href{#PureStateBox}{Pure States and Star Products}, and
\href{#SHOBox}{Simple Harmonic Oscillator}).

QMPS can obviously shed light on subtle quantization problems as the
comparison with classical theories is more systematic and natural. \ Since the
variables involved are the same in both classical and quantum cases, the
connection to the classical limit as $\hbar\rightarrow0$ is more readily
apparent (see \href{#ClassicalLimitBox}{Classical Limit}). \ But beyond this
and self-evident pedagogical intuition, what is this alternate formulation of
QM and its panoply of satisfying mathematical structures good for?

It is the natural language to describe quantum transport, and to monitor
decoherence of macroscopic quantum states, in interaction with the
environment, a pressing central concern of quantum computing \cite{Preskill}.
\ It can\ also serve to analyze and quantize physics phenomena unfolding in an
hypothesized \emph{noncommutative spacetime }with various noncommutative
geometries \cite{Szabo}. \ Such phenomena are most naturally described in
Groenewold's and Moyal's language.

However, it may be fair to say that, as was true for the path integral
formulation during the first few decades of its existence, the best QMPS
\textquotedblleft killer apps\textquotedblright\ are yet to come.%
%TCIMACRO{\FRAME{dtbpFU}{2.8561in}{2.0587in}{0pt}{\Qcb{Work by CoBrA artist
%Mogens Balle (1921-1988).}}{}{killerapps.eps}%
%width 2.8561in;  height 2.0587in;  depth 0pt;  original-width 3.9695in;
%original-height 2.8522in;  cropleft "0";  croptop "1";  cropright "1";
%cropbottom "0";  filename 'KillerApps.eps';file-properties "XNPEU";}} }%
%BeginExpansion
\begin{center}
\includegraphics[
height=2.0587in,
width=2.8561in
]%
{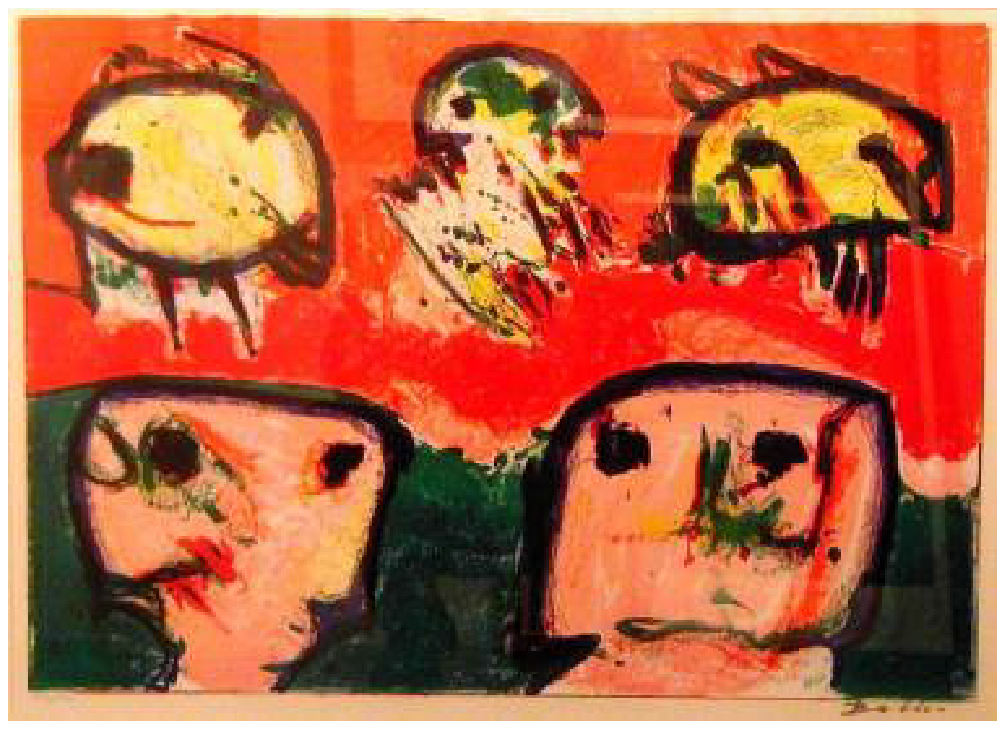}%
\\
Work by CoBrA artist Mogens Balle (1921-1988).
\end{center}
%EndExpansion

\subsubsection*{Acknowledgement}

\textit{We thank David Fairlie for many insightful discussions about QMPS.
\ We also thank Carolyne van Vliet for providing information about Dutch
physicists. \ This work was supported in part by NSF Award 0855386, and in
part by the U.S. Department of Energy, Division of High Energy Physics, under
contract DE-AC02-06CH11357.}

\section*{Appendices}

The following supplements expand on the discussion in the text. \ The first
three are historical commentaries, including biographical material, while the
remaining supplements deal with elementary technical aspects of QMPS.\bigskip

\noindent\fbox{\hypertarget{DiracBox}{Dirac}}\hrulefill\bigskip

A representative, indeed \emph{authoritative}, opinion, dismissing even the
suggestion that quantum mechanics can be expressed in terms of
classical-valued phase space variables, was expressed by Paul Dirac in a
letter to Joe Moyal on 20 April 1945 (see p 135, \cite{AnnMoyal}). \ Dirac
said, \textquotedblleft I think it is obvious that there cannot be any
distribution function $F\left(  p,q\right)  $ which would give correctly the
mean value of any $f\left(  p,q\right)  $ ...\textquotedblright\ \ He then
tried to carefully explain why he thought as he did, by discussing the
underpinnings of the uncertainty relation.

However, in this instance, Dirac's opinion was wrong, and unfounded, despite
the fact that he must have been thinking about the subject since\ publishing
some preliminary work along these lines many years before \cite{Dirac1930}.
\ In retrospect, it is Dirac's unusual misreading of the situation that is
obvious, rather than the non-existence of $F\left(  p,q\right)  $. \ 

Perhaps the real irony here is that Dirac's brother-in-law, Eugene Wigner, had
already constructed such an $F\left(  p,q\right)  $ several years earlier
\cite{Wigner}. \ Moyal eventually learned of Wigner's work and brought it to
Dirac's attention in a letter dated 21 August 1945 (see p 159, \cite{AnnMoyal}%
). \ 

Nevertheless, the historical record strongly suggests that Dirac held fast to
his opinion that quantum mechanics could \emph{not} be formulated in terms of
classical-valued phase space variables. \ For example, Dirac made no changes
when discussing the von Neumann density operator, $\boldsymbol{\rho}$, on p
132 in the final edition of his book \cite{Dirac}. \ Dirac maintained
\textquotedblleft Its existence is rather surprising in view of the fact that
phase space has no meaning in quantum mechanics, there being no possibility of
assigning numerical values simultaneously to the $q$'s and $p$%
's.\textquotedblright\ \ This statement completely overlooks the fact that the
Wigner function $F\left(  p,q\right)  $ is precisely a realization of
$\boldsymbol{\rho}$ in terms of numerical-valued $q$'s and $p$'s.

But how could it be, with his unrivaled ability to create elegant theoretical
physics, Dirac did \emph{not} seize the opportunity so unmistakably laid
before him, by Moyal, to return to his very first contributions to the theory
of quantum mechanics and examine in greater depth the relation between
classical Poisson brackets and quantum commutators? \ We will probably never
know beyond any doubt --- yet another sort of uncertainty principle --- but we
are led to wonder if it had to do with some key features of Moyal's theory at
that time. \ First, in sharp contrast to Dirac's own operator methods, in its
initial stages QMPS theory was definitely \emph{not} a pretty formalism!
\ And, as is well known, beauty was one of Dirac's guiding principles in
theoretical physics. \ 

Moreover, the logic of the early formalism was not easy to penetrate. \ It is
clear from his correspondence with Moyal that Dirac did not succeed in cutting
away the formal undergrowth to clear a precise conceptual path through the
theory behind QMPS, or at least not one that \emph{he} was eager to travel
again.\footnote{Although Dirac did pursue closely related ideas, at least once
\cite{Dirac1945}, in his contribution to Bohr's festschrift.}

One of the main reasons the early formalism was not pleasing to the eye, and
nearly impenetrable, may have had to do with another key aspect of Moyal's
1945 theory: \ Two constructs may have been missing. \ Again, while we cannot
be absolutely certain, we \emph{suspect} the star product and the related
bracket were both absent from Moyal's theory \emph{at that time}. \ So far as
we can tell, neither of these constructs appears in any of the correspondence
between Moyal and Dirac.\marginpar{%
%TCIMACRO{\FRAME{itbpFU}{1.5108in}{3.4938in}{0in}{\Qcb{P Dirac}}{}%
%{diracaxe.eps}{\special{ language "Scientific Word";  type "GRAPHIC";
%width 1.5108in;  height 3.4938in;  depth 0in;  original-width 2.3843in;
%original-height 5.6533in;  cropleft "0";  croptop "1";  cropright "1";
%cropbottom "0";  filename 'DiracAxe.eps';file-properties "XNPEU";}} }%
%BeginExpansion
{\parbox[b]{1.5108in}{\begin{center}
\includegraphics[
height=3.4938in,
width=1.5108in
]%
{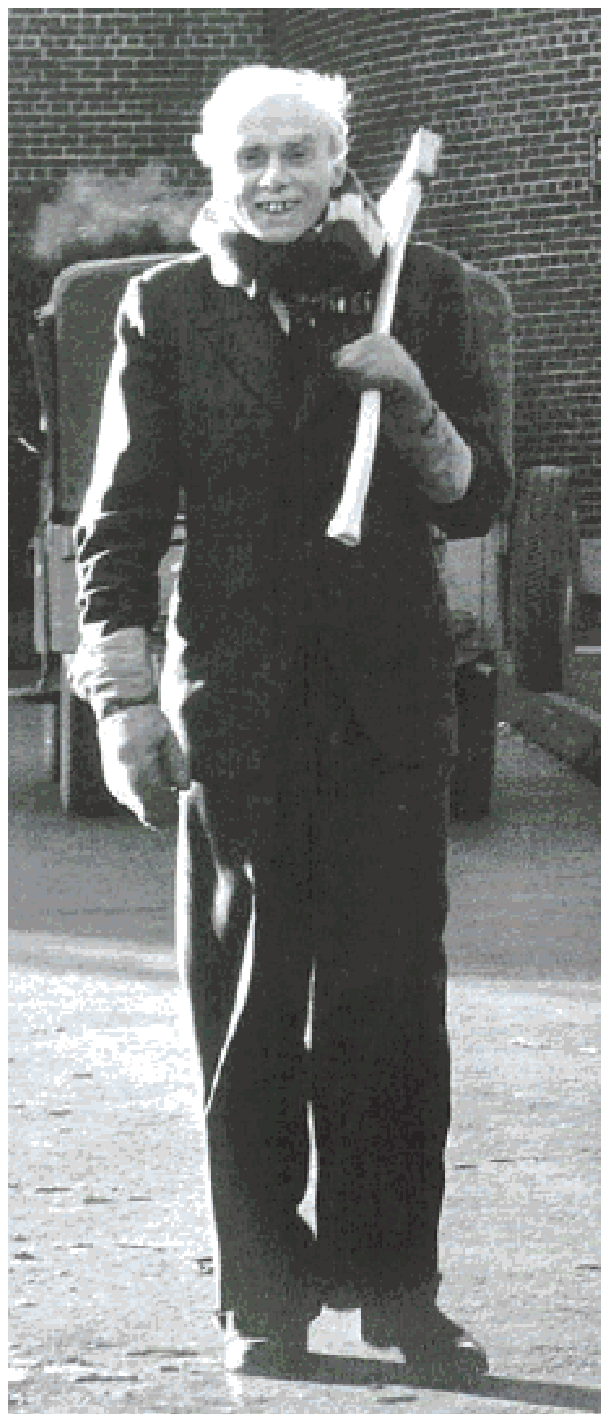}%
\\
P Dirac
\end{center}}}
%EndExpansion
} \ In fact, the product itself is not even contained in the published form of
Moyal's work that appeared four years later \cite{Moyal}, although the
antisymmetrized version of the product --- the so-called Moyal bracket --- is
articulated in that work as a generalization of the Poisson
bracket,\footnote{See Eqn (7.10) and the associated comments in the last
paragraph of \S 7, p 106, \cite{Moyal}.} after first being used by Moyal to
express the time evolution of $F\left(  p,q;t\right)  $.\footnote{See Eqn
(7.8), \cite{Moyal}. \ Granted, the equivalent of that equation was already
available in \cite{Wigner}, but Wigner did \emph{not} make the sweeping
generalization offered by Moyal's Eqn (7.10).} \ Even so, we are not aware of
any historical evidence that Moyal \emph{specifically} brought his bracket to
Dirac's attention.

Thus, we can hardly avoid speculating, had Moyal communicated \emph{only the
contents of his single paragraph about the generalized bracket}%
\footnotemark[6] to Dirac, the latter would have recognized its importance, as
well as its beauty, and the discussion between the two men would have acquired
an altogether different tone. \ For, as Dirac wrote to Moyal on 31 October
1945 (see p 160, \cite{AnnMoyal}), \textquotedblleft I think your kind of work
would be valuable only if you can put it in a very neat
form.\textquotedblright\ \ The Groenewold product and the Moyal bracket do
just that.\footnote{In any case, by then Groenewold had already found the star
product, as well as the related bracket, by taking Weyl's and von Neumann's
ideas to their logical conclusion, and had it all published \cite{Groenewold}
in the time between Moyal's and Dirac's last correspondence and the appearance
of \cite{Moyal,BartlettMoyal}, wherein discussions with Groenewold are
acknowledged by Moyal.}\newpage

\noindent\fbox{\hypertarget{GroenewoldBox}{Groenewold}}\hrulefill
\marginpar{\bigskip
\par%
%TCIMACRO{\FRAME{itbpFU}{1.4062in}{1.8334in}{0in}{\Qcb{H Groenewold}}%
%{}{groenewold.eps}{\special{ language "Scientific Word";  type "GRAPHIC";
%width 1.4062in;  height 1.8334in;  depth 0in;  original-width 8.2027in;
%original-height 10.7539in;  cropleft "0";  croptop "1";  cropright "1";
%cropbottom "0";  filename 'Groenewold.eps';file-properties "XNPEU";}} }%
%BeginExpansion
{\parbox[b]{1.4062in}{\begin{center}
\includegraphics[
height=1.8334in,
width=1.4062in
]%
{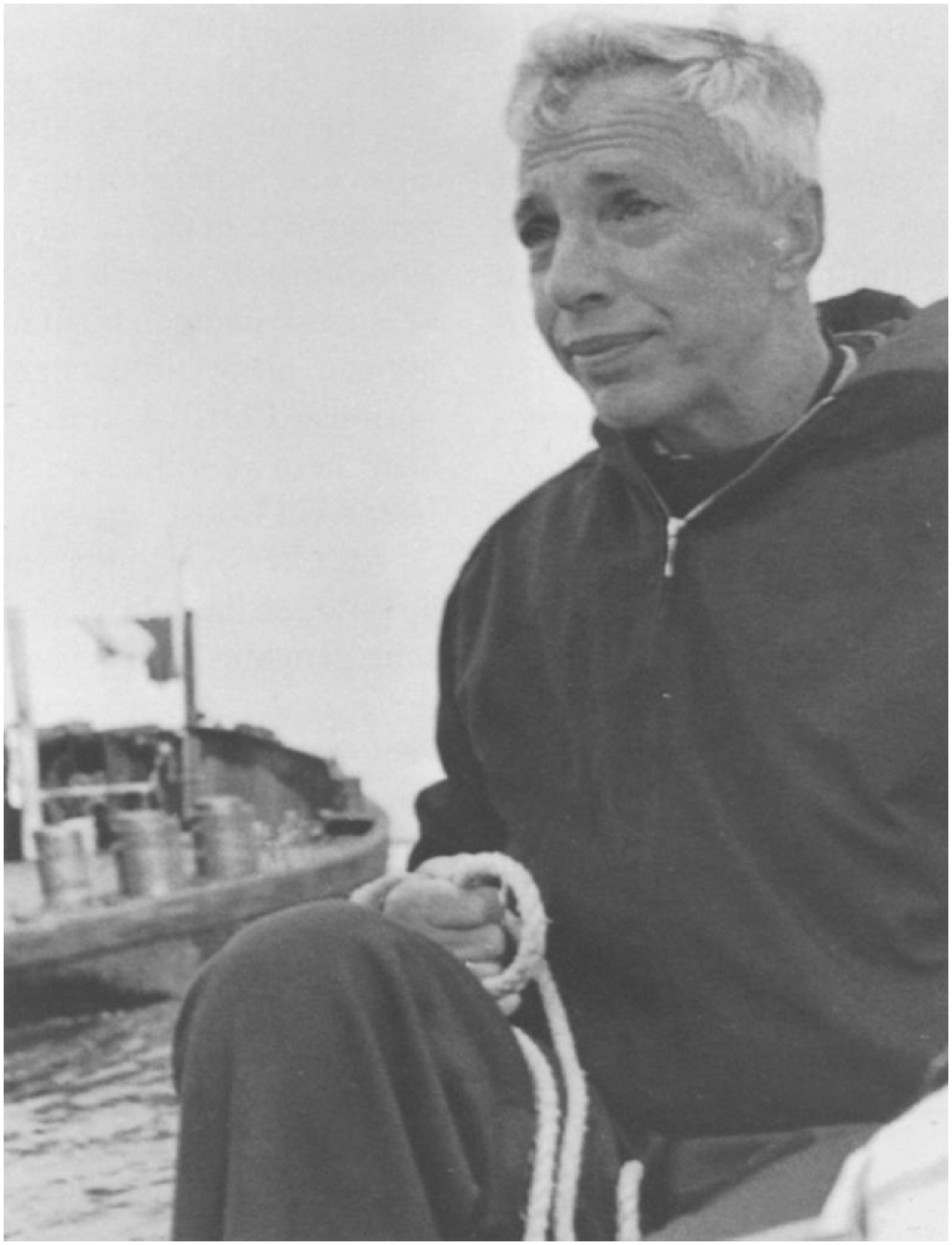}%
\\
H Groenewold
\end{center}}}
%EndExpansion
}

\subsection*{Hilbrand Johannes Groenewold}

29 June 1910 - 23 November 1996\footnote{The material presented here contains
statements taken from a previously published obituary \cite{Hugenholtz}.}

Hip Groenewold was born in Muntendam, The Netherlands. \ He studied at the
University of Groningen, from which he graduated in physics with subsidiaries
in mathematics and mechanics in 1934.

In that same year, he went of his own accord to Cambridge, drawn by the
presence there of the mathematician John von Neumann, who had given a solid
mathematical foundation to quantum mechanics with his book
\textit{Mathematische Grundlagen der Quantenmechanik}. \ This period had a
decisive influence on Groenewold's scientific thinking. \ During his entire
life, he remained especially interested in the interpretation of quantum
mechanics (e.g. some of his ideas are recounted in \cite{ManyWorlds}). \ It is
therefore not surprising that his PhD thesis, which he completed eleven years
later, was devoted to this subject \cite{Groenewold}. \ In addition to his
revelation of the star product, and associated technical details, Groenewold's
achievement in his thesis was to escape the cognitive straight-jacket of the
mainstream view that the defining difference between classical mechanics and
quantum mechanics was the use of c-number functions and operators,
respectively.\ \ He understood that these were only habits of use and in no
way restricted the physics.

Ever since his return from England in 1935 until his permanent appointment at
theoretical physics in Groningen in 1951, Groenewold experienced difficulties
finding a paid job in physics. \ He was an assistant to Zernike in Groningen
for a few years, then he went to the Kamerlingh Onnes Laboratory in Leiden,
and taught at a grammar school in the Hague from 1940 to 1942. \ There, he met
the woman whom he married in 1942. \ He spent the remaining war years at
several locations in the north of the Netherlands. \ In July 1945, he began
work for another two years as an assistant to Zernike. \ Finally, he worked
for four years at the KNMI (Royal Dutch Meteorological Institute) in De Bilt.

During all these years, Groenewold never lost sight of his research. \ At his
suggestion upon completing his PhD thesis, in 1946, Rosenfeld, of the
University of Utrecht, became his promoter, rather than Zernike.\ \ In 1951,
he was offered a position at Groningen in theoretical physics: \ First as a
lecturer, then as a senior lecturer, and finally as a professor in 1955.
\ With his arrival at the University of Groningen, quantum mechanics was
introduced into the curriculum. \ 

In 1971 he decided to resign as a professor in theoretical physics in order to
accept a position in the Central Interfaculty for teaching Science and
Society. \ However, he remained affiliated with the theoretical institute as
an extraordinary professor. \ In 1975 he retired.

In his younger years, Hip was a passionate puppet player, having brought
happiness to many children's hearts with beautiful puppets he made himself.
\ Later, he was especially interested in painting. \ He personally knew
several painters, and owned many of their works. \ He was a great lover of the
after-war CoBrA art. \ This love gave him much comfort during his last years.

\hrulefill\newpage

\noindent\fbox{\hypertarget{MoyalBox}{Moyal}}\hrulefill\marginpar{{}
\par
\bigskip
\par%
%TCIMACRO{\FRAME{itbpFU}{1.407in}{2.0375in}{0in}{\Qcb{J Moyal}}{}%
%{moyal.eps}{\special{ language "Scientific Word";  type "GRAPHIC";
%width 1.407in;  height 2.0375in;  depth 0in;  original-width 2.0678in;
%original-height 3.0199in;  cropleft "0";  croptop "1";  cropright "1";
%cropbottom "0";  filename 'Moyal.eps';file-properties "XNPEU";}} }%
%BeginExpansion
{\parbox[b]{1.407in}{\begin{center}
\includegraphics[
height=2.0375in,
width=1.407in
]%
{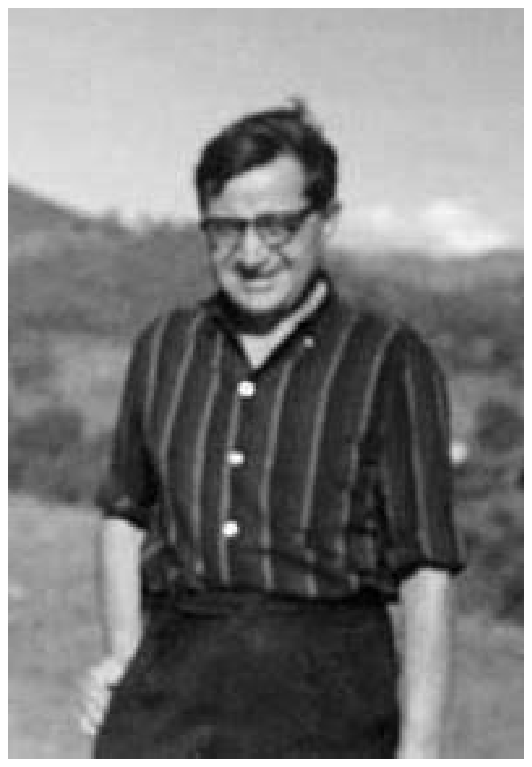}%
\\
J Moyal
\end{center}}}
%EndExpansion
}

\subsection*{Jos\'{e} Enrique Moyal}

1 October 1910 - 22 May 1998\footnote{The material presented here contains
statements taken from a previously published obituary \cite{Gani}.}

Joe Moyal was born in Jerusalem and spent much of his youth in Palestine. \ He
studied electrical engineering in France, at Grenoble and Paris, in the early
1930s. He then worked as an engineer, later continuing his studies in
mathematics at Cambridge, statistics at the Institut de Statistique, Paris,
and theoretical physics at the Institut Henri Poincar\'{e}, Paris.

After a period of research on turbulence and diffusion of gases at the French
Ministry of Aviation in Paris, he escaped to London at the time of the German
invasion in 1940. \ The eminent writer C.P. Snow, then adviser to the British
Civil Service, arranged for him to be allocated to de Havilland's at Hatfield,
where he was involved in aircraft research into vibration and electronic instrumentation.

During the war, hoping for a career in theoretical physics, Moyal developed
his ideas on the statistical nature of quantum mechanics, initially trying to
get Dirac interested in them, in December 1940, but without success. \ After
substantial progress on his own, his poignant and intense scholarly
correspondence with Dirac (Feb 1944 to Jan 1946, reproduced in \cite{AnnMoyal}%
) indicates he was not aware, at first, that his phase-space statistics-based
formulation was actually equivalent to standard QM. \ Nevertheless, he soon
appreciated its alternate beauty and power. \ In their spirited
correspondence, Dirac patiently but insistently recorded his reservations,
with mathematically trenchant arguments, although lacking essential
appreciation of Moyal's novel point of view: \ A radical departure from the
conventional Hilbert space picture \cite{Moyal}. \ The correspondence ended in
anticipation of a Moyal colloquium at Cambridge in early 1946.

That same year, Moyal's first academic appointment was in Mathematical Physics
at Queen's University Belfast. \ He was later a lecturer and senior lecturer
with M.S. Bartlett in the Statistical Laboratory at the University of
Manchester, where he honed and applied his version of quantum mechanics
\cite{BartlettMoyal}.

In 1958, he became a Reader in the Department of Statistics, Institute of
Advanced Studies, Australian National University, for a period of 6 years.
\ There he trained several graduate students, now eminent professors in
Australia and the USA. \ In 1964, he returned to his earlier interest in
mathematical physics at the Argonne National Laboratory near Chicago, coming
back to Macquarie University as Professor of Mathematics before retiring in 1978.

Joe's interests were broad: \ He was an engineer who contributed to the
understanding of rubber-like materials; a statistician responsible for the
early development of the mathematical theory of stochastic processes; a
theoretical physicist who discovered the \textquotedblleft Moyal
bracket\textquotedblright\ in quantum mechanics; and a mathematician who
researched the foundations of quantum field theory. \ He was one of a rare
breed of mathematical scientists working in several fields, to each of which
he made fundamental contributions.

\hrulefill\newpage

\noindent\fbox{\hypertarget{StarProductsBox}{Star Product}}\hrulefill

The \emph{star product} is the Weyl correspondent of the Hilbert space
operator product, and was developed through the work of many over a number of
years: \ H Weyl (1927), J von Neumann (1931), E Wigner (1932), H Groenewold
(1946), J Moyal (1949), and\ G Baker (1958), as well as more recent work to
construct the product on general manifolds (reprinted in \cite{QMPS}, along
with related papers). \ There are useful integral and differential
realizations of the product:%
\[
f\star g=\int\frac{dx_{1}dp_{1}}{2\pi\left(  \hbar/2\right)  }\int\frac
{dx_{2}dp_{2}}{2\pi\left(  \hbar/2\right)  }\;f\left(  x+x_{1},p+p_{1}\right)
\;g\left(  x+x_{2},p+p_{2}\right)  \;\exp\left(  \frac{i}{\hbar/2}\left(
x_{1}p_{2}-x_{2}p_{1}\right)  \right)  \ ,
\]%
\[
x_{1}p_{2}-x_{2}p_{1}=\text{Area}\left(  \text{1,2}\;\text{parallelogram}%
\right)  \ ,\ \ \ \hbar/2=\text{Planck Area}=\min\left(  \Delta x\Delta
p\right)  \ ,
\]%
\begin{align*}
& \\
f\star g  &  =f\left(  x,p\right)  \;\exp\left(  \overleftarrow{\partial_{x}%
}\,\frac{i\hbar}{2}\,\overrightarrow{\partial_{p}}-\overleftarrow{\partial
_{p}}\,\frac{i\hbar}{2}\,\overrightarrow{\partial_{x}}\right)  \;g\left(
x,p\right)  \ ,
\end{align*}%
\begin{align*}
f\star g  &  =f\left(  x+\frac{1}{2}i\hbar\,\overrightarrow{\partial_{p}%
}\;,\;p-\frac{1}{2}i\hbar\,\overrightarrow{\partial_{x}}\right)  \;g\left(
x,p\right) \\
&  =f\left(  x,p\right)  \;g\left(  x-\frac{1}{2}i\hbar\overleftarrow{\partial
_{p}}\;,\;p+\frac{1}{2}i\hbar\,\overleftarrow{\partial_{x}}\right) \\
&  =f\left(  x+\frac{1}{2}i\hbar\,\overrightarrow{\partial_{p}}\;,\;p\right)
\;g\left(  x-\frac{1}{2}i\hbar\overleftarrow{\partial_{p}}\;,\;p\right) \\
&  =f\left(  x,p-\frac{1}{2}i\hbar\,\overrightarrow{\partial_{x}}\right)
\;g\left(  x,p+\frac{1}{2}i\hbar\,\overleftarrow{\partial_{x}}\right)  \ .
\end{align*}
\marginpar{%
%TCIMACRO{\FRAME{itbpFU}{0.8994in}{1.4935in}{0in}{\Qcb{H Weyl}}{}%
%{weyl1930.eps}{\special{ language "Scientific Word";  type "GRAPHIC";
%width 0.8994in;  height 1.4935in;  depth 0in;  original-width 3.7334in;
%original-height 6.2759in;  cropleft "0";  croptop "1";  cropright "1";
%cropbottom "0";  filename 'Weyl1930.eps';file-properties "XNPEU";}} }%
%BeginExpansion
{\parbox[b]{0.8994in}{\begin{center}
\includegraphics[
height=1.4935in,
width=0.8994in
]%
{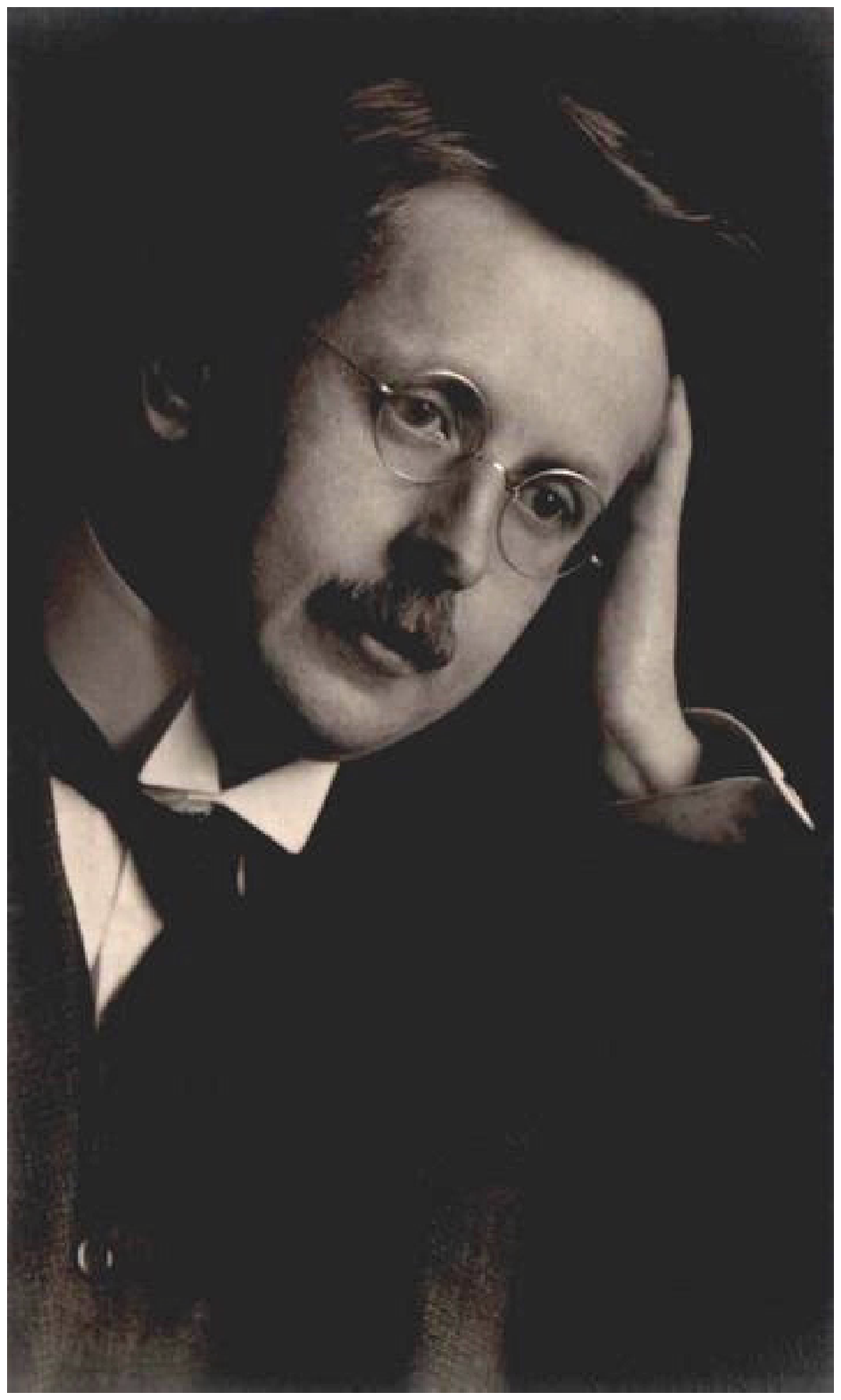}%
\\
H Weyl
\end{center}}}
%EndExpansion
}The \emph{Moyal bracket}, $\left\{  \!\left\{  f,g\right\}  \!\right\}
=\frac{1}{i\hbar}\left[  f,g\right]  _{\star}$, is essentially just the
antisymmetric part of a star product,\ where%
\[
\left[  f,g\right]  _{\star}\overset{\text{defn.}}{=}f\star g-g\star f\ .
\]
This provides a homomorphism with commutators of operators, e.g. $\left[
x,p\right]  _{\star}=i\hbar$.

\hrulefill\hrulefill\newpage

\noindent\fbox{\hypertarget{WFBox}{Wigner Functions}}\hrulefill

Wigner's original definition of his eponymous function\ was (Eqn(5) in
\cite{Wigner})%
\begin{multline*}
\mathcal{P}\left(  x_{1},\cdots,x_{n};p_{1},\cdots,p_{n}\right)  =\left(
\frac{1}{\hbar\pi}\right)  ^{n}\int_{-\infty}^{\infty}\cdots\int dy_{1}\cdots
dy_{n}\psi\left(  x_{1}+y_{1}\cdots x_{n}+y_{n}\right)  ^{\ast}\\
\psi\left(  x_{1}-y_{1}\cdots x_{n}-y_{n}\right)  e^{2i\left(  p_{1}%
y_{1}+\cdots+p_{n}y_{n}\right)  /\hbar}\ .
\end{multline*}
So defined, Wigner functions (WFs)\emph{\ reside in phase space}.

WFs are the Weyl-correspondents of von Neumann's density operators,
$\mathbf{\rho}$. \ Thus, in terms of Hilbert space position and momentum
\emph{operators} $\boldsymbol{X}$ and $\boldsymbol{P}$, we
have\footnote{Remark on units: \ As defined by Wigner, WFs have units of
$1/\hbar^{n}$ in $2n$-dimensional phase space. \ Since it is customary for the
density operator\ to have no units, a compensating factor of $\hbar^{n}$ is
required in the Weyl correspondence relating WFs to $\boldsymbol{\rho}$s.
Issues about units are most easily dealt with if one works in
\textquotedblleft action-balanced\textquotedblright\ $x$ and $p$ variables,
whose units are $\left[  x\right]  =\left[  p\right]  =\sqrt{\hbar}$.}
\[
\mathbf{\rho}=\frac{\hbar^{n}}{\left(  2\pi\right)  ^{2n}}\int d^{n}\tau
d^{n}\sigma d^{n}xd^{n}p\,\mathcal{P}\left(  x_{1},\cdots,x_{n};p_{1}%
,\cdots,p_{n}\right)  \,\exp\left(  i\tau\cdot\left(  \boldsymbol{P}-p\right)
+i\sigma\cdot\left(  \boldsymbol{X}-x\right)  \right)  \ .
\]

In one $x$ and one $p$ dimension, denoting the WF by $F\left(  x,p\right)  $
instead of $\mathcal{P}$ (in deference to the momentum operator
$\boldsymbol{P}$), we have
\begin{align*}
F\left(  x,p\right)   &  =\frac{1}{\pi\hbar}\int dy\,\left\langle
x+y\right\vert \,\mathbf{\rho\,}\left\vert x-y\right\rangle \,e^{-2ipy/\hbar
}\ ,\\
\left\langle x+y\right\vert \,\mathbf{\rho\,}\left\vert x-y\right\rangle  &
=\int dp\,F\left(  x,p\right)  \,e^{2ipy/\hbar}\ ,\\
\mathbf{\rho}  &  =2\int dxdy\int dp\,\left\vert x+y\right\rangle \,F\left(
x,p\right)  \,e^{2ipy/\hbar}\,\left\langle x-y\right\vert \ .
\end{align*}
For a quantum mechanical \textquotedblleft pure state\textquotedblright%
\begin{align*}
F\left(  x,p\right)   &  =\frac{1}{\pi\hbar}\int dy\,\psi\left(  x+y\right)
\,\psi^{\ast}\left(  x-y\right)  \,e^{-2ipy/\hbar}\ ,\\
\psi\left(  x+y\right)  \,\psi^{\ast}\left(  x-y\right)   &  =\int
dp\,F\left(  x,p\right)  \,e^{2ipy/\hbar}\ ,\\
\mathbf{\rho}  &  \mathbf{=}\left\vert \psi\right\rangle \left\langle
\psi\right\vert \ ,
\end{align*}
where as usual, $\psi\left(  x+y\right)  =\left\langle x+y\right.  \left\vert
\psi\right\rangle ,\;\left\langle \psi\right\vert \left.  x-y\right\rangle
=\psi^{\ast}\left(  x-y\right)  $. \ Direct application of the
Cauchy--Bunyakovsky--Schwarz inequality to the first of these pure-state
relations gives%
\[
\left\vert F\left(  x,p\right)  \right\vert \leq\frac{1}{\pi\hbar}\int
dy\,\left\vert \psi\left(  y\right)  \right\vert ^{2}\ .
\]
So, for normalized states with $\int dy\,\left\vert \psi\left(  y\right)
\right\vert ^{2}=1$, we have the bounds%
\[
-\frac{1}{\pi\hbar}\leq F\left(  x,p\right)  \leq\frac{1}{\pi\hbar}\ .
\]
Such normalized states therefore cannot give probability spikes (e.g. Dirac
deltas) without taking the classical limit $\hbar\rightarrow0$. \ The
corresponding bound in $2n$ phase-space dimensions is given by the same
argument applied to Wigner's original definition: \ $\left\vert \mathcal{P}%
\left(  x_{1},\cdots,x_{n};p_{1},\cdots,p_{n}\right)  \right\vert \leq\left(
\frac{1}{\hbar\pi}\right)  ^{n}$.

\hrulefill\marginpar{%
%TCIMACRO{\FRAME{itbpFU}{1.0084in}{1.4157in}{0in}{\Qcb{E Wigner}}{}%
%{wigner.eps}{\special{ language "Scientific Word";  type "GRAPHIC";
%width 1.0084in;  height 1.4157in;  depth 0in;  original-width 0.7057in;
%original-height 1.0023in;  cropleft "0";  croptop "1";  cropright "1";
%cropbottom "0";  filename 'Wigner.eps';file-properties "XNPEU";}} }%
%BeginExpansion
{\parbox[b]{1.0084in}{\begin{center}
\includegraphics[
height=1.4157in,
width=1.0084in
]%
{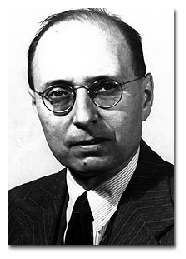}%
\\
E Wigner
\end{center}}}
%EndExpansion
}\newpage

\noindent\fbox{\hypertarget{PureStateBox}{Pure States and Star Products}%
}\hrulefill

\emph{Pure-state} Wigner functions must obey a \emph{projection} condition.
\ If the normalization is set to the standard value%
\[
\iint_{-\infty}^{+\infty}dxdp\,F\left(  x,p\right)  =1\ ,
\]
then the function corresponds to a pure state if and only if \
\[
F=\left(  2\pi\hbar\right)  \,F\star F\ .
\]
These statements correspond to the pure-state density operator conditions:
\ $Tr\left(  \mathbf{\rho}\right)  =1$ and $\mathbf{\rho}=\mathbf{\rho\,\rho}%
$, respectively. \ 

\bigskip

If both of the above are true, then $F$ describes an allowable pure state for
a quantized system. \ Otherwise not. \ You can easily satisy only one out of
these two conditions, but not the other, using an $F$ that is not a pure state.

\bigskip

Without drawing on the Hilbert space formulation, it may at first seem to be
rather remarkable that explicit WFs actually satisfy the projection condition
(cf. the above Gaussian example, for the only situation where it works,
$a=b=1$, i.e. $\exp\left(  -\left(  x^{2}+p^{2}\right)  /\hbar\right)  $).
\ However, if $F$ is known to be a $\star$ eigenfunction with non-vanishing
eigenvalue of some phase-space function with a \emph{non}-degenerate spectrum
of eigenvalues, then it must be true that $F\varpropto F\star F$ as a
consequence of associativity, since both $F$\ and $F\star F$ would yield the
same eigenvalue.

\hrulefill\hrulefill

\bigskip

\noindent\fbox{\hypertarget{ExercisesBox}{Exercises}}\hrulefill

Some simple exercises for students to sharpen their QMPS skills. \ Show the following.

\begin{exercise}
Non-commutativity.
\begin{align*}
e^{ax+bp}\star e^{Ax+Bp}  &  =e^{\left(  a+A\right)  x+\left(  b+B\right)
p}\;e^{\left(  aB-bA\right)  i\hbar/2}\\
&  \neq\\
e^{Ax+Bp}\star e^{ax+bp}  &  =e^{\left(  a+A\right)  x+\left(  b+B\right)
p}\;e^{\left(  Ab-Ba\right)  i\hbar/2}%
\end{align*}

\end{exercise}

\begin{exercise}
Associativity.
\[
\left(  e^{ax+bp}\star e^{Ax+Bp}\right)  \star e^{\alpha x+\beta p}%
\]%
\[
=e^{\left(  a+A+\alpha\right)  x+\left(  b+B+\beta\right)  p}\;e^{\left(
aB-bA+a\beta\mathcal{-}b\alpha+A\beta\mathcal{-}B\alpha\right)  i\hbar/2}%
\]%
\[
=e^{ax+bp}\star\left(  e^{Ax+Bp}\star e^{\alpha x+\beta p}\right)
\]

\end{exercise}

\begin{exercise}
Trace properties. \ \ \ \ (a.k.a. \textquotedblleft Lone Star
Lemma\textquotedblright)
\[
\int dxdp\,f\star g=\int dxdp\,f\,g=\int dxdp\,g\,f=\int dxdp\,g\star f
\]

\end{exercise}

\begin{exercise}
Gaussians. \ \ \ \ For $a,b\geq0$,%
\[
\exp\left(  -\frac{a}{\hbar}\left(  x^{2}+p^{2}\right)  \right)  ~\star
~\exp\left(  -\frac{b}{\hbar}\left(  x^{2}+p^{2}\right)  \right)  =\frac
{1}{1+ab}\exp\left(  -\frac{a+b}{\left(  1+ab\right)  \hbar}\left(
x^{2}+p^{2}\right)  \right)  \;.
\]

\end{exercise}

Additional exercises may be culled from the first chapter of \cite{QMPS}.

\hrulefill\hrulefill\newpage

\noindent\fbox{\hypertarget{SHOBox}{The Simple Harmonic Oscillator}}%
\hrulefill\marginpar{%
%TCIMACRO{\FRAME{itbpFU}{2.0055in}{1.6034in}{0in}{\Qcb{$n=0$}}{}{sho0.eps}%
%width 2.0055in;  height 1.6034in;  depth 0in;  original-width 4.8049in;
%original-height 3.8277in;  cropleft "0";  croptop "1";  cropright "1";
%cropbottom "0";  filename 'SHO0.eps';file-properties "XNPEU";}} }%
%BeginExpansion
{\parbox[b]{2.0055in}{\begin{center}
\includegraphics[
height=1.6034in,
width=2.0055in
]%
{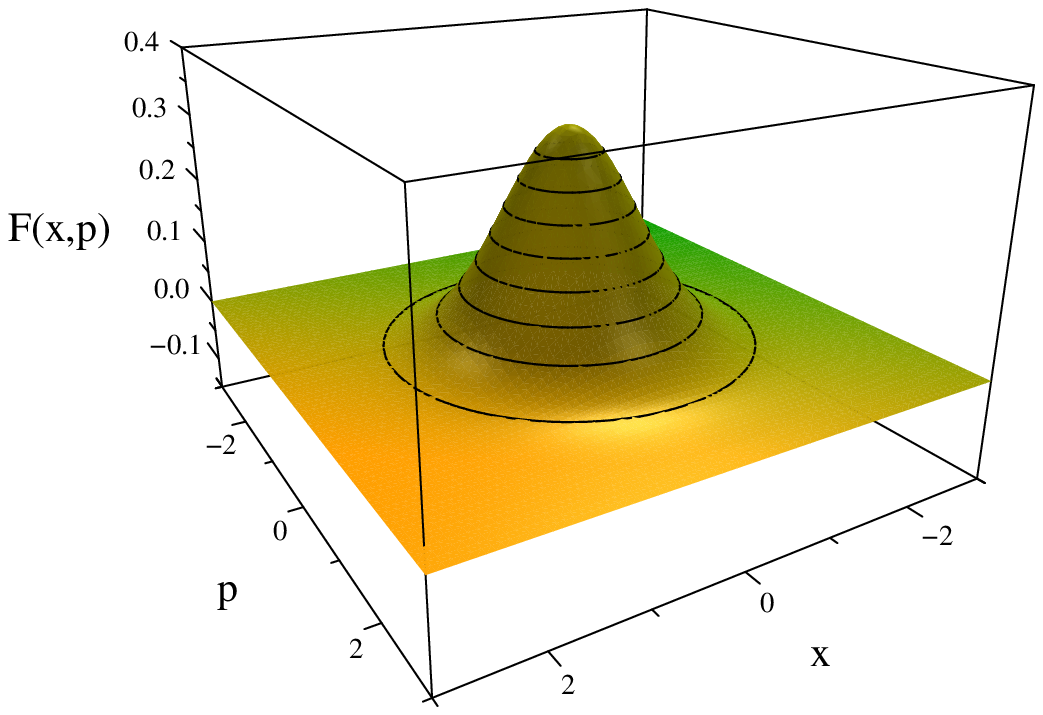}%
\\
$n=0$%
\end{center}}}
%EndExpansion
\par%
%TCIMACRO{\FRAME{itbpFU}{2.0055in}{1.6034in}{0in}{\Qcb{$n=1$}}{}{sho1.eps}%
%width 2.0055in;  height 1.6034in;  depth 0in;  original-width 4.8049in;
%original-height 3.8277in;  cropleft "0";  croptop "1";  cropright "1";
%cropbottom "0";  filename 'SHO1.eps';file-properties "XNPEU";}} }%
%BeginExpansion
{\parbox[b]{2.0055in}{\begin{center}
\includegraphics[
height=1.6034in,
width=2.0055in
]%
{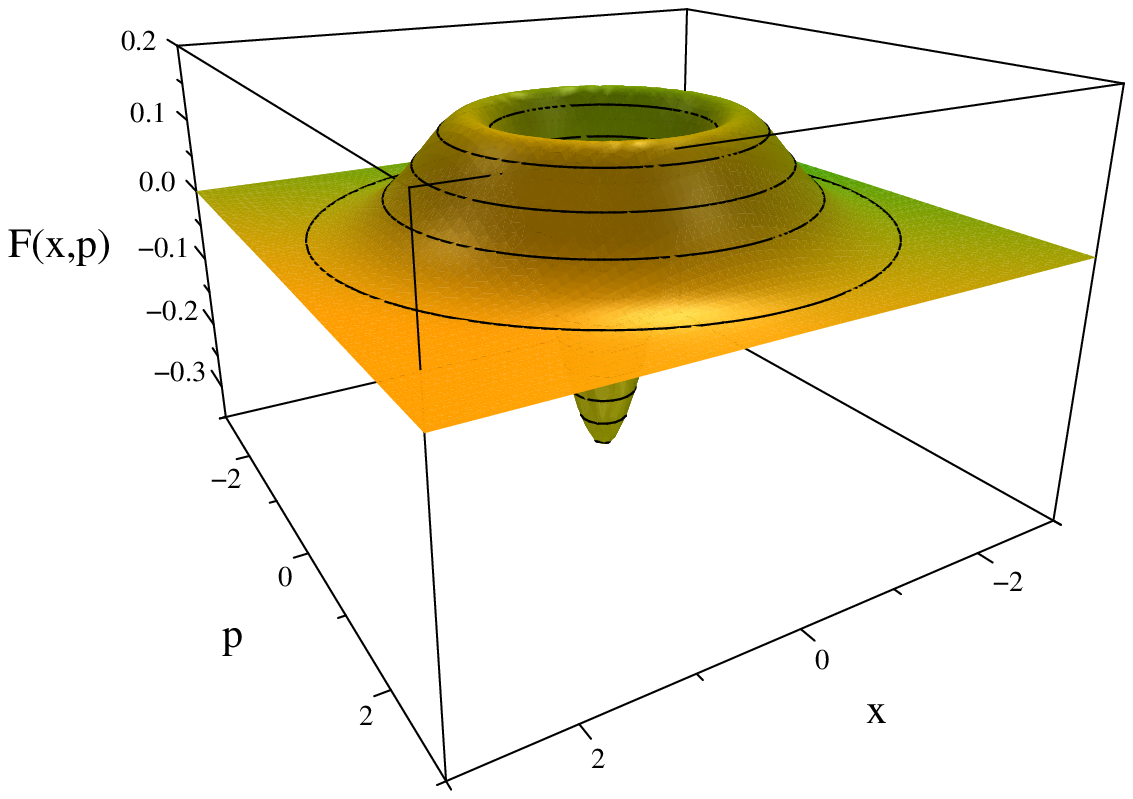}%
\\
$n=1$%
\end{center}}}
%EndExpansion
\par%
%TCIMACRO{\FRAME{itbpFU}{2.0055in}{1.6025in}{0in}{\Qcb{$n=2$}}{}{sho2.eps}%
%width 2.0055in;  height 1.6025in;  depth 0in;  original-width 4.8049in;
%original-height 3.8277in;  cropleft "0";  croptop "1";  cropright "1";
%cropbottom "0";  filename 'SHO2.eps';file-properties "XNPEU";}} }%
%BeginExpansion
{\parbox[b]{2.0055in}{\begin{center}
\includegraphics[
height=1.6025in,
width=2.0055in
]%
{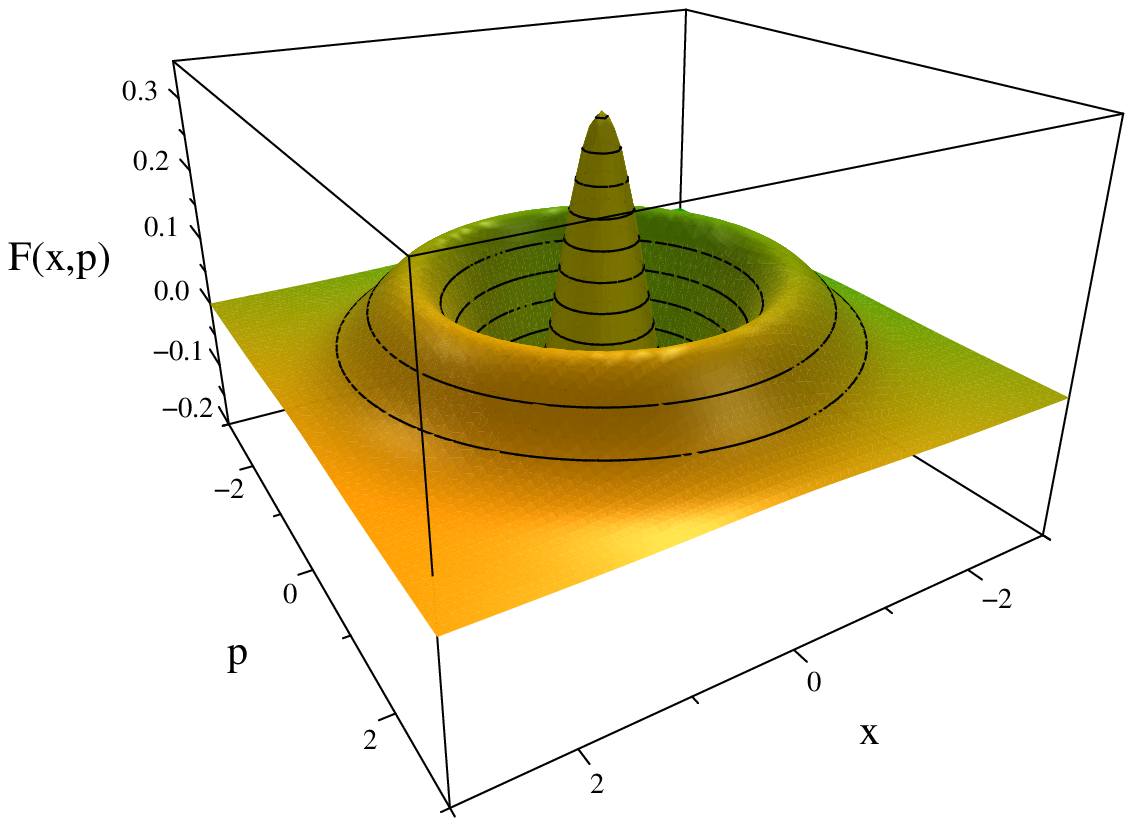}%
\\
$n=2$%
\end{center}}}
%EndExpansion
\par%
%TCIMACRO{\FRAME{itbpFU}{2.0055in}{1.6025in}{0in}{\Qcb{$n=3$}}{}{sho3.eps}%
%width 2.0055in;  height 1.6025in;  depth 0in;  original-width 4.8049in;
%original-height 3.8277in;  cropleft "0";  croptop "1";  cropright "1";
%cropbottom "0";  filename 'SHO3.eps';file-properties "XNPEU";}} }%
%BeginExpansion
{\parbox[b]{2.0055in}{\begin{center}
\includegraphics[
height=1.6025in,
width=2.0055in
]%
{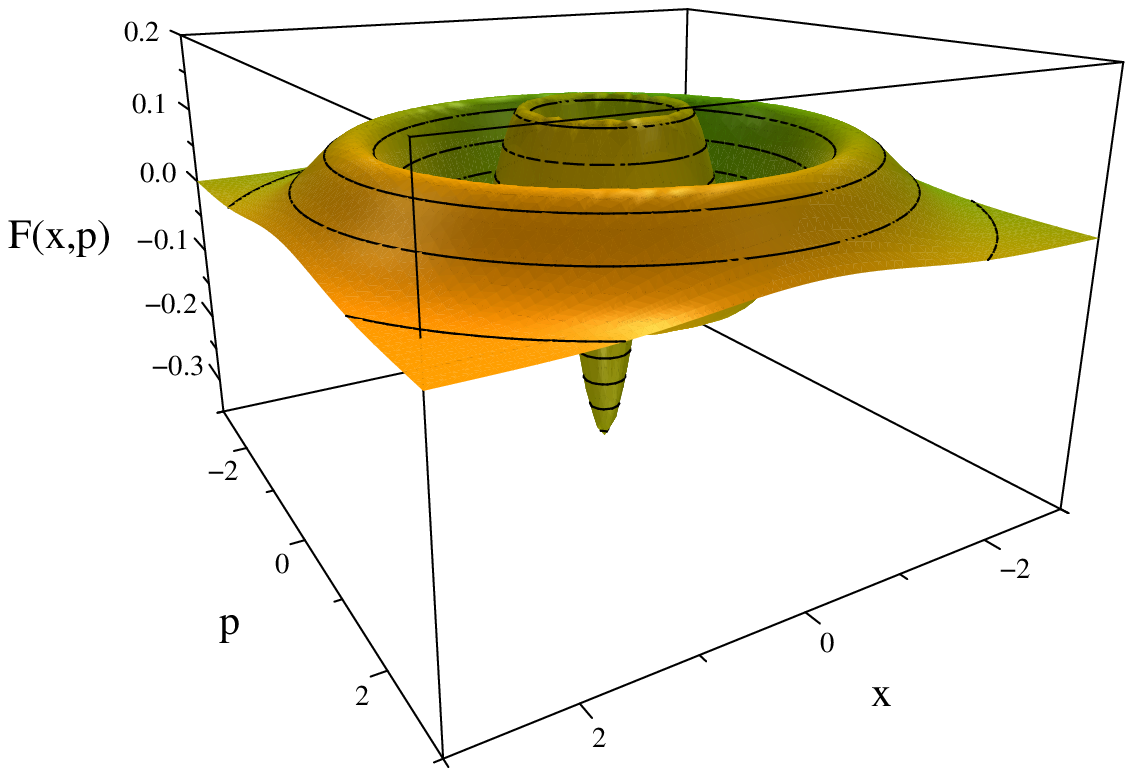}%
\\
$n=3$%
\end{center}}}
%EndExpansion
}

There is no need to deal with wave functions or Hilbert space states. \ The
WFs may be constructed directly on the phase space
\cite{Groenewold,BartlettMoyal}. \ Energy eigenstates are obtained as (real)
solutions of the $\star$\emph{-genvalue} equations \cite{Fairlie,QMPS}:%

\[
H\star F=E\,F=F\star H\ .
\]
To illustrate this, consider the simple harmonic oscillator (SHO) with ($m=1$,
$\omega=1$)
\[
H=\frac{1}{2}\left(  p^{2}+x^{2}\right)  \ .
\]
The above equations are now second-order \emph{partial} differential
equations,%
\begin{align*}
H\star F  &  =\frac{1}{2}\left(  \left(  p-\frac{1}{2}i\hbar\,\partial
_{x}\right)  ^{2}+\left(  x+\frac{1}{2}i\hbar\,\partial_{p}\right)
^{2}\right)  F=EF\ ,\\
F\star H  &  =\frac{1}{2}\left(  \left(  p+\frac{1}{2}i\hbar\,\partial
_{x}\right)  ^{2}+\left(  x-\frac{1}{2}i\hbar\,\partial_{p}\right)
^{2}\right)  F=EF\ .
\end{align*}
But if we subtract (or take the imaginary part),%
\[
\left(  p\partial_{x}-x\partial_{p}\right)  F=0\;\;\;\;\;\Rightarrow
\;\;\;\;\;F\left(  x,p\right)  =F\left(  x^{2}+p^{2}\right)  \ .
\]
So $H\star F=EF=F\star H$ reduces to a single \emph{ordinary} differential
equation\ (Laguerre, not Hermite!), namely, the real part of either of the
previous second-order equations. \ 

There are integrable solutions if and only if $E=\left(  n+1/2\right)  \hbar$,
$n=0,1,\cdots$ for which%
\begin{align*}
F_{n}\left(  x,p\right)   &  =\frac{\left(  -1\right)  ^{n}}{\pi\hbar}%
L_{n}\left(  \frac{x^{2}+p^{2}}{\hbar/2}\right)  \,e^{-\left(  x^{2}%
+p^{2}\right)  /\hbar}\ ,\\
L_{n}\left(  z\right)   &  =\frac{1}{n!}e^{z}\frac{d^{n}}{dz^{n}}\left(
z^{n}e^{-z}\right)  \ .
\end{align*}
The normalization is chosen to be the standard one $\iint_{-\infty}^{+\infty
}dxdp\,F_{n}\left(  x,p\right)  =1$. \ Except for the $n=0$ ground state
(Gaussian) WF, these $F$'s change sign on the $xp$-plane. For example:
$L_{0}\left(  z\right)  =1,\;L_{1}\left(  z\right)  =1-z,\;L_{2}\left(
z\right)  =1-2z+\frac{1}{2}z^{2}$, etc.

Using the integral form of the $\star$ product, it is now easy to check these
pure states are $\star$ orthogonal:
\[
\left(  2\pi\hbar\right)  \,F_{n}\star F_{k}=\delta_{nk}\,F_{n}\ .
\]
This becomes more transparent by using $\star$ raising/lowering operations to
write\footnote{Note that the earlier exercise giving the star composition law
of Gaussians immediately yields the projection property of the SHO ground
state WF, $F_{0}=\left(  2\pi\hbar\right)  F_{0}\star F_{0}$.}%
\begin{align*}
F_{n}  &  =\frac{1}{n!}\left(  a^{\ast}\star\right)  ^{n}\,F_{0}\,\left(
\star a\right)  ^{n}\\
&  =\frac{1}{\pi\hbar n!}\left(  a^{\ast}\star\right)  ^{n}\,e^{-\left(
x^{2}+p^{2}\right)  /\hbar}\,\left(  \star a\right)  ^{n}\ ,
\end{align*}
where $a$ is the usual linear combination$\;a\equiv\frac{1}{\sqrt{2\hbar}%
}(x+ip),\;$and $a^{\ast}$ is just its complex conjugate $a^{\ast}\equiv
\frac{1}{\sqrt{2\hbar}}(x-ip)$ , with $a\star a^{\ast}-a^{\ast}\star a=1,$ and
$a\star F_{0}=0=F_{0}\star a^{\ast}$ (cf. coherent state density operators).

\hrulefill\hrulefill\newpage\noindent
\fbox{\hypertarget{UncertaintyPrincipleBox}{The Uncertainty Principle}%
}\hrulefill

\emph{Expectation values} of all phase-space functions, say $G\left(
x,p\right)  $, for a system described by $F\left(  x,p\right)  $ (a real WF,
normalized to $1$) \emph{are just integrals} of \emph{ordinary} products (cf.
Lone Star Lemma)
\[
\left\langle G\right\rangle =\int dxdp\,G\left(  x,p\right)  \,F\left(
x,p\right)  \ .
\]
These can be negative, even though $G$ is positive, if the WF flips sign. \ So
how do we directly establish simple correct statements such as $\left\langle
\left(  x+p\right)  ^{2}\right\rangle \geq0$ without using marginal
probabilities or invoking Hilbert space results?

The roles of positive-definite Hilbert space operators are played on phase
space by \emph{real star-squares}:
\[
G\left(  x,p\right)  =g^{\ast}\left(  x,p\right)  \star g\left(  x,p\right)
\ .
\]
These \emph{always} have non-negative expectation values for any $g$ and any
WF,%
\[
\langle g^{\ast}\star g\rangle\geq0\ ,
\]
even if $F$ becomes negative. \ (Note this is \emph{not} true if the $\star$
is removed! $\ $If $g^{\ast}\star g$ is supplanted by $\left\vert g\right\vert
^{2}$, then integrated with a WF, the result could be negative.)

To show this, first suppose the system is in a pure state. \ Then use
$F=\left(  2\pi\hbar\right)  \,F\star F$ (see \href{#PureStateBox}{Pure States
and Star Products}), and the associativity and trace properties (see
\href{#ExercisesBox}{Exercises}), to write{\small :\footnote{{\small By
essentially the same argument, if $F_{1}$ and $F_{2}$ are two distinct pure
state WFs, the phase-space overlap integral between the two is also manifestly
non-negative and thus admits interpretation as the transition probability
between the respective states:
\[
\int dxdp\,F_{1}F_{2}=({2\pi\hbar})^{2}\int dxdp\,|F_{1}\star F_{2}|^{2}%
\geq0\ .
\]
}}}%
\begin{align*}
\int dxdp\;\left(  g^{\ast}\star g\right)  \;F  &  =\left(  2\pi\hbar\right)
\int dxdp\ \left(  g^{\ast}\star g\right)  \left(  F\star F\right) \\
&  =\left(  2\pi\hbar\right)  \int dxdp\ \left(  g^{\ast}\star g\right)
\star\left(  F\star F\right) \\
&  =\left(  2\pi\hbar\right)  \int dxdp\ \left(  g^{\ast}\star g\star
F\right)  \star F\\
&  =\left(  2\pi\hbar\right)  \int dxdp\ F\star\left(  g^{\ast}\star g\star
F\right) \\
&  =\left(  2\pi\hbar\right)  \int dxdp\ \left(  F\star g^{\ast}\right)
\star\left(  g\star F\right) \\
&  =\left(  2\pi\hbar\right)  \int dxdp\ \left(  F\star g^{\ast}\right)
\left(  g\star F\right) \\
&  =\left(  2\pi\hbar\right)  \int dxdp\ |g\star F|^{2}\\
&  \geq0\ .
\end{align*}
In the next to last step we also used the elementary property $(g\star
F)^{\ast}=F\star g^{\ast}$. \ 

More generally, if the system is in a mixed state, as defined by a normalized
probabilistic sum of pure states, $F=\sum_{j}\mathcal{P}_{j}F_{j}$, with
probabilities $\mathcal{P}_{j}\geq0$ satisfying $\sum_{j}\mathcal{P}_{j}=1$,
then the same inequality holds.\newpage

Correlations of observables follow conventionally from specific choices of
$g(x,p)$. \ For example, to produce Heisenberg's uncertainty relation, take%
\[
g\left(  x,p\right)  =a+bx+cp\ ,
\]
for arbitrary complex coefficients $a,b,c$. \ The resulting positive
semi-definite quadratic form is then%
\[
\langle g^{\ast}\star g\rangle=\left(
\begin{array}
[c]{ccc}%
a^{\ast} & b^{\ast} & c^{\ast}%
\end{array}
\right)  \left(
\begin{array}
[c]{ccc}%
1 & \langle x\rangle & \langle p\rangle\\
\langle x\rangle & \langle x\star x\rangle & \langle x\star p\rangle\\
\langle p\rangle & \langle p\star x\rangle & \langle p\star p\rangle
\end{array}
\right)  \left(
\begin{array}
[c]{c}%
a\\
b\\
c
\end{array}
\right)  \geq0\ ,
\]
for any $a,b,c$. \ All eigenvalues of the above $3\times3$ hermitian matrix
are therefore non-negative, and thus so is its determinant,%
\[
\det\left(
\begin{array}
[c]{ccc}%
1 & \langle x\rangle & \langle p\rangle\\
\langle x\rangle & \langle x\star x\rangle & \langle x\star p\rangle\\
\langle p\rangle & \langle p\star x\rangle & \langle p\star p\rangle
\end{array}
\right)  \geq0\ .
\]
But
\[
x\star x=x^{2}\;,\qquad p\star p=p^{2}\;,\qquad x\star p=xp+i\hbar/2\;,\qquad
p\star x=xp-i\hbar/2\;,
\]
and with the usual definitions of the variances
\[
(\Delta x)^{2}\equiv\langle(x-\langle x\rangle)^{2}\rangle\;,\qquad(\Delta
p)^{2}\equiv\langle(p-\langle p\rangle)^{2}\rangle\;,
\]
the positivity condition on the above determinant amounts to
\[
(\Delta x)^{2}(\Delta p)^{2}\geq\frac{1}{4}\hbar^{2}+\left(  \left\langle
xp\right\rangle -\langle x\rangle\langle p\rangle\right)  ^{2}\;.
\]
Hence Heisenberg's relation
\[
\left(  \Delta x\right)  \left(  \Delta p\right)  \geq\hbar/2\ .
\]
The inequality is saturated for a vanishing original integrand $g\star F=0$,
for suitable $a,b,c$, when the $\left\langle xp\right\rangle -\langle
x\rangle\langle p\rangle$ term vanishes (i.e. $x$ and $p$ statistically
independent, as happens for a Gaussian pure state, $F=\frac{1}{\pi\hbar}%
\exp\left(  -\left(  x^{2}+p^{2}\right)  /\hbar\right)  $).

\hrulefill\marginpar{%
%TCIMACRO{\FRAME{itbpFU}{1.1113in}{1.4935in}{0in}{\Qcb{W Heisenberg}}%
%{}{heisenberg.eps}{\special{ language "Scientific Word";  type "GRAPHIC";
%width 1.1113in;  height 1.4935in;  depth 0in;  original-width 3.3624in;
%original-height 4.5506in;  cropleft "0";  croptop "1";  cropright "1";
%cropbottom "0";  filename 'Heisenberg.eps';file-properties "XNPEU";}} }%
%BeginExpansion
{\parbox[b]{1.1113in}{\begin{center}
\includegraphics[
height=1.4935in,
width=1.1113in
]%
{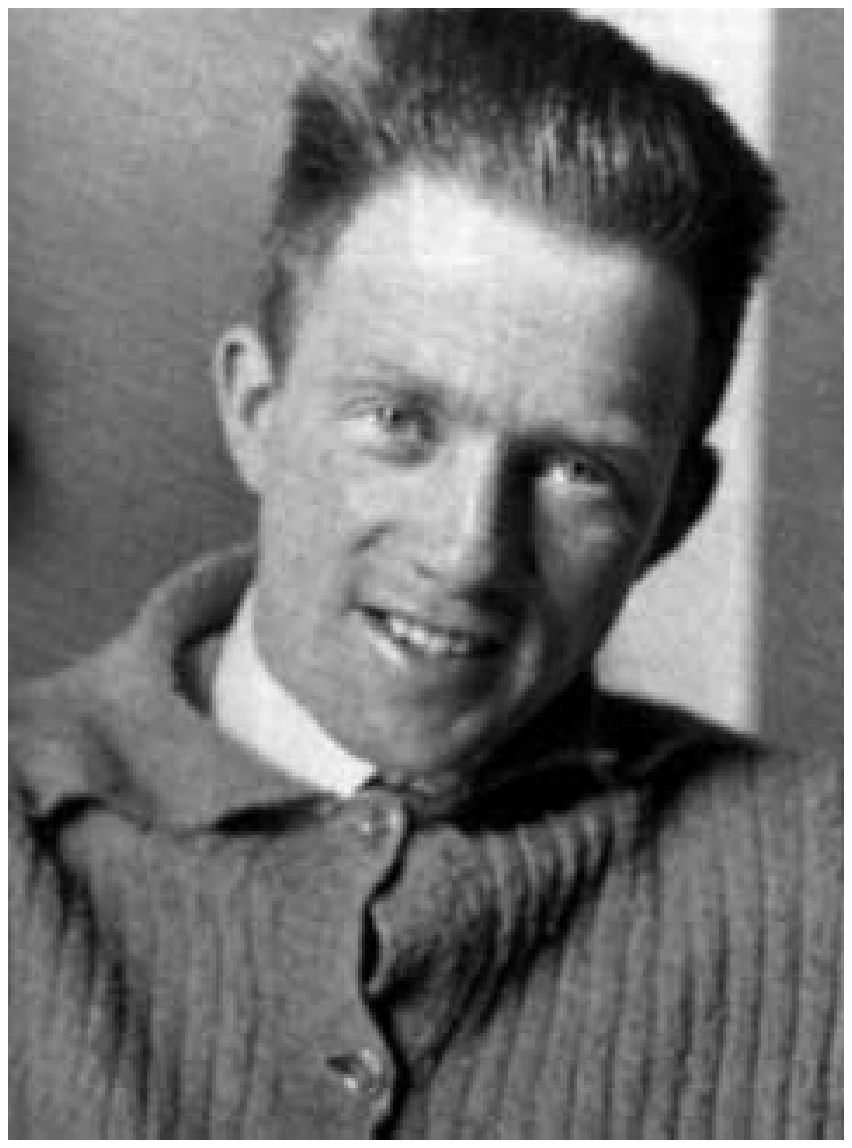}%
\\
W Heisenberg
\end{center}}}
%EndExpansion
}

\bigskip

\noindent\fbox{\hypertarget{ClassicalLimitBox}{A Classical Limit}}\hrulefill

The simplest illustration of the classical limit, \emph{by far}, is provided
by the SHO ground state WF. \ In the limit $\hbar\rightarrow0$ a completely
localized phase-space distribution is obtained, namely, a Dirac delta at the
origin of the phase space:
\[
\lim_{\hbar\rightarrow0}\frac{1}{\pi\hbar}\exp\left(  -\left(  x^{2}%
+p^{2}\right)  /\hbar\right)  =\delta\left(  x\right)  \delta\left(  p\right)
\ .
\]
Moreover, if the ground state Gaussian is uniformly displaced from the origin
by an amount $\left(  x_{0},p_{0}\right)  $ and then allowed to evolve in
time, its peak follows a classical trajectory (see
\href{http://server.physics.miami.edu/curtright/TimeDependentWignerFunctions.html}{Movies}
... this simple behavior does \emph{not} hold for less trivial potentials).
\ The classical limit of this evolving WF is therefore just a Dirac delta
whose spike follows the trajectory of a classical point particle moving in the
harmonic potential:%
\begin{multline*}
\lim_{\hbar\rightarrow0}\frac{1}{\pi\hbar}\exp\left(  -\left(  \left(
x-x_{0}\cos t-p_{0}\sin t\right)  ^{2}+\left(  p-p_{0}\cos t+x_{0}\sin
t\right)  ^{2}\right)  /\hbar\right) \\
=\delta\left(  x-x_{0}\cos t-p_{0}\sin t\right)  \delta\left(  p-p_{0}\cos
t+x_{0}\sin t\right)  \ .
\end{multline*}

\hrulefill

\hrulefill
\end{document}